\documentclass[aps]{revtex4}
\usepackage[dvips]{epsfig}
\usepackage{amsmath,amssymb}

\parskip=\medskipamount

\newcommand{\eq}[1]{(\ref{#1})}
\newcommand{\fig}[1]{Fig.\ref{#1}}

\newcommand{\be}{\begin{equation}}
\newcommand{\ee}{\end{equation}}

\newcommand{\barr}{\begin{array}}
\newcommand{\earr}{\end{array}}

\newcommand{\beqn}{\begin{eqnarray}}
\newcommand{\eeqn}{\end{eqnarray}}

\newcommand{\bs}{\begin{subequations}}
\newcommand{\es}{\end{subequations}}

\newcommand{\bw}{\begin{widetext}}
\newcommand{\ew}{\end{widetext}}

\newcommand\disp{\displaystyle}

\newcommand{\la}{\left<}
\newcommand{\ra}{\right>}

\newcommand{\up}{\uparrow}
\newcommand{\down}{\downarrow}
\newcommand{\lef}{\leftarrow}
\newcommand{\rig}{\rightarrow}

\newcommand{\Peak}{\operatorname{Peak}}

\newcommand{\card}{\operatorname{\#}}

\begin{document}

\title{On the distribution of surface extrema in several one-- and two--dimensional
random landscapes}
\author{F. Hivert$^{1}$, S.Nechaev$^{2,4}$, G.Oshanin$^{3,}$\footnote{Also at:
Max-Planck-Institut f\"ur Metallforschung, Heisenbergstr. 3, D-70569 Stuttgart,
Germany, and Institut f\"ur Theoretische und Angewandte Physik, Universit\"at
Stuttgart, Pfaffenwaldring 57, D-70569 Stuttgart, Germany},
O.Vasilyev$^{3,4,}$\footnote{Present address: Center for Molecular Modelling,
Materia Nova, University of Mons-Hainaut, Mons, Belgium}}

\affiliation{$^{1}$Institut Gaspard Monge, Universit\'e de Marne-la-Vall\'ee 5,
77454 Marne la Vall\'ee Cedex 2, France \\ $^{2}$LPTMS, Universit\'e Paris Sud,
91405 Orsay Cedex, France \\ $^{3}$LPTMC, Universit\'e Paris 6, 4 Place Jussieu,
75252 Paris, France \\ $^4$L.D. Landau Institute for Theoretical Physics, 117334,
Moscow, Russia}

\date{May 18, 2006}

\begin{abstract}

We study here a standard next-nearest-neighbor (NNN) model of ballistic growth on
one- and two-dimensional substrates focusing our analysis on the probability
distribution function $P(M,L)$ of the number $M$ of maximal points (i.e., local
``peaks'') of growing surfaces. Our analysis is based on two central results: (i)
the proof (presented here) of the fact that uniform one--dimensional ballistic
growth process in the steady state can be mapped onto ''rise-and-descent'' sequences
in the ensemble of random permutation matrices; and (ii) the fact, established in
Ref. \cite{ov}, that different characteristics of ``rise-and-descent'' patterns in
random permutations can be interpreted in terms of a certain continuous--space
Hammersley--type process. For one--dimensional system we compute $P(M,L)$ exactly
and also present explicit results for the correlation function characterizing the
enveloping surface. For surfaces grown on 2d substrates, we pursue similar approach
considering the ensemble of permutation matrices with long--ranged correlations.
Determining exactly the first three cumulants of the corresponding distribution
function, we define it in the scaling limit using an expansion in the Edgeworth
series, and show that it converges to a Gaussian function as $L \to \infty$.

\end{abstract}

\maketitle

\section{Introduction}
\label{sect:1}

Within the recent years much effort has been devoted to theoretical analysis of
properties of surfaces obtained by aggregation of particles. Several models
describing various properties of clusters grown by different deposition processes
have been proposed. To name but a few, we mention the famous Kardar--Parisi--Zhang
(KPZ) \cite{KPZ} and the Edwards--Wilkinson (EW) models \cite{EW}, models of
surfaces grown by Molecular Beam Epitaxy (MBE) (see, for example, Ref.\cite{MBE}),
Polynuclear Growth (PNG) \cite{M,PS,BR,J} and by Ballistic Deposition (BD)
\cite{Mand,MRSB,KM,BMW,KTKR}, in which case particles are sequentially added to a
growing surface along ballistic trajectories with random initial positions and
specified direction.

For these models of surface growth, a number of important theoretical achievements
concerning the statistics of extrema has been made. In particular, in Ref.\cite{SC}
the distributions of the maximal heights of the 1D Edwards--Wilkinson and of the KPZ
interfaces were determined exactly. In Ref.\cite{PS} it was realized that the height
distribution of the PNG surfaces coincides with the so-called Tracy--Widom
distribution \cite{TW}, which appears in the theory of random matrices. In
Ref.\cite{scaling} it has been shown that in the thermodynamic limit BD exhibits the
KPZ scaling behavior. Moreover, a discrete BD model has been shown recently to be a
very convenient tool for studying non-Abelian entanglement properties of braided
directed random walks \cite{nech}. Finally, it has been found that in many models of
ballistic growth, as well as in their continuum--space counterparts belonging to the
KPZ universality class \cite{KPZ}, the average velocity of cluster's growth is
governed by the density of local minima of the enveloping surface \cite{KTKR}.

In this work we analyze the structure of the enveloping surface in a standard
next-nearest-neighbor (NNN) model of ballistic growth on one-- and two--dimensional
substrates. Interpreting this model in terms of permutations of the set $1,2,3,
\ldots, L$, where $L$ is the number of lattice sites and the numbers drawn at random
from this set determine local heights of the surface,  we calculate the Probability
Distribution Function (PDF) of the number of maximal points (i.e., local ``peaks'').
Our analysis is based on two central results:

(i) A proof, presented in this paper, of the fact that a ballistic growth process in
the steady state can be formulated exactly in terms of a ``rise--and--descent''
pattern in the ensemble of random permutation matrices;

(ii) An observation made in Ref. \cite{ov} that the ``rise--and--descent'' patterns
can be treated efficiently using a recently proposed algorithm of a Permutation
Generated Random Walk.

We  remark that the expected value and the variance of the number of local peaks in
surfaces grown by ballistic deposition on a one-dimensional and on a two-dimensional
honeycomb lattices have been already calculated by J. Desbois in Ref.\cite{des}
using a certain decoupling of the hierarchy of coupled differential equations
describing evolution of the moments of higher order. This method provides correct
results for the first two moments of the distribution function and, apparently, may
be extended further for the calculation of higher moments by truncation of the
higher level correlations. However, this approach is not completely rigorous. Our
approach, on contrary, is mathematically exact, does not rely on any uncontrollable
assumption and enables us to go beyond the results obtained in \cite{des}. In
particular, in one dimension we calculate a complete distribution function of the
number of local peaks exactly. Apart of that, we also present explicit results for
the correlation functions characterizing the enveloping surface. For surfaces
emerging in two--dimensional ballistic growth, we reduce the problem to the analysis
of the ensemble of permutation matrices with long--ranged correlations. Determining
exactly the first three cumulants of the corresponding PDF, we obtain the
distribution function in the scaling limit using expansions in the Edgeworth series.

The paper is outlined as follows. In Section II we formulate our models in one-- and
two--dimensions, discuss, on an intuitive level, a relation between a sequential
growth of patterns in ballistic aggregation and "dynamics" on permutations, and
finally present the main results of this work. Section III is devoted to a rigorous
description of the relation between ballistic deposition and an "updating" dynamics
on permutations. Further on, in Section IV we analyze the probability distribution
function of the number of local peaks on one-dimensional substrates. Here, we
briefly outline the well-known results of Stembridge on the peak numbers
\cite{stem}, describe the model and basic results obtained for the so-called
Permutation Generated Random Walks \cite{ov} and show how the operator formalism
developed in this work can be extended for the calculation of the moments of the
probability distribution function. Next, in Section V we present a derivation of an
explicit expression of the probability of having two local surface peaks at distance
$l$ apart of each other (such that there is no peaks inbetween). In Section VI we
focus on two-dimensional situation and show how our previous analysis can be
extended to this case. We calculate exactly first three cumulants of the
corresponding probability distribution function of having $M$ peaks on a square
lattice containing $L$ sites and then, using an expansion in the Edgeworth series,
show that this function converges to a Gaussian as $L \to \infty$. Finally, in
Section VII we conclude with a brief summary of our results.

\section{Models, Definitions and Main Results}
\label{sect:2}

\subsection{Surfaces in standard ballistic growth process}

A standard one--dimensional ballistic deposition model with
next--nearest--neighboring (NNN) interactions is formulated as follows (for more
details, see Ref.\cite{scaling}). Consider a box divided in $L$ columns (of unit
width each) enumerated by index $i$ ($i=1,2,...,L$). For simplicity, we assume the
periodic boundary conditions, such that the leftmost and the rightmost columns are
neighbors.

\begin{figure}[ht]
\epsfig{file=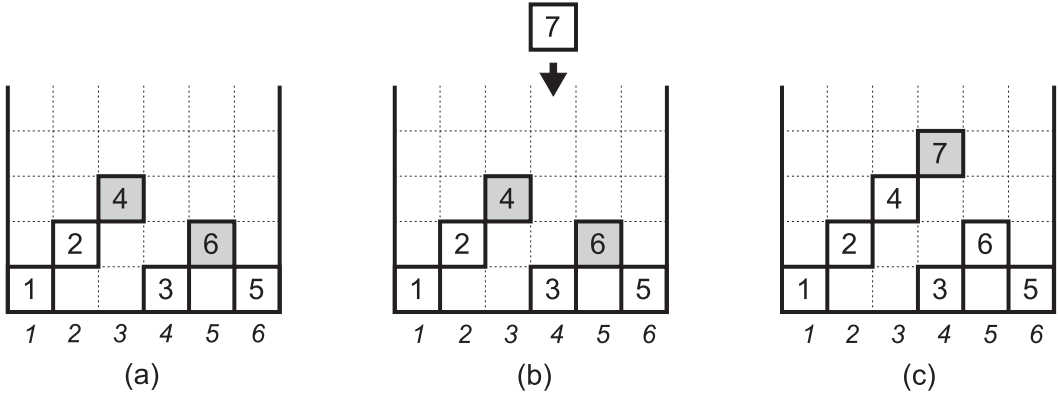, width=14cm} \caption{Sequential growth of a heap in the
ballistic deposition process.} \label{fig:1a}
\end{figure}

At the initial time moment, ($n=0$), the system is deemed empty. Then, at each tick
of the clock, $n=1,2,...,N$,  we deposit an elementary cell (``particle'') of unit
height and width in a randomly chosen column. Suppose that the distribution on the
set of columns is uniform. Define the height, $h(i,n)$, in the column $i$ at time
moment $n$. Assume now, as it is depicted in \fig{fig:1a}, that the cells in the
nearest--neighboring columns interact in such a way that they can only touch each
other by corners, but never by their vertical sides. This implies that after having
deposited a particle to the column $i$, the height of this column is modified
according to the following rule: \be h(i,n+1)=\max\{h(i-1,n),\, h(i,n),\,
h(i+1,n)\}+1 \label{eq:1}. \ee If at the time moment $n$ nothing is added to the
column $i$, its height remains unchanged: $h(i,n+1)=h(i,n)$. A set of deposited
particles forms a pile as shown in \fig{fig:1a}a for $L=6$ columns and $N=6$
particles. Here, for example, $h(1,6)=1$, $h(2,6)=2$, etc.

\begin{figure}[ht]
\epsfig{file=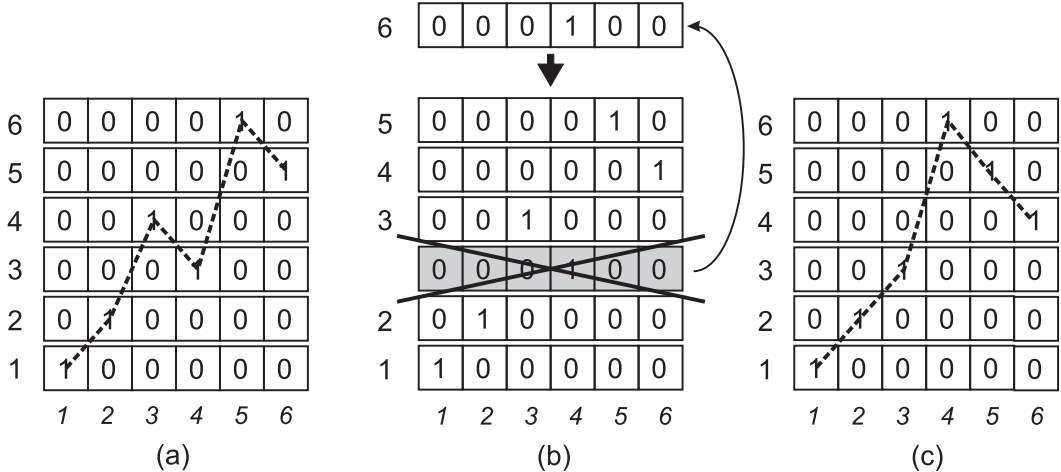, width=14cm} \caption{An "updating dynamics" on
permutations.} \label{fig:1b}
\end{figure}

Now, we call  the local maxima of the pile as the ``peaks''. More specifically, take
the set $\cal{H}$ of heights at some time moment $n$: ${\cal H}=\{h(1,n), h(2,n),
..., h(L,n)\}$. We say that the column $i$ contains a "peak" at time moment $n$ if
the height of this column satisfies the following two--sided inequality: \be
\left\{\begin{array}{l} h(i,n)>h(i-1,n) \medskip \\ h(i,n)>h(i+1,n).
\end{array}\right. \label{eq:2}
\ee
Note that in \fig{fig:1a} there are two peaks situated in the columns $i=3$ and
$i=5$. The collection of peaks ${\cal T}$ is the subset of ${\cal H}$ and forms the
``roof'' --- the set of upmost (or ``removable'' \footnote{Only the particle of the
roof ${\cal T}$ can be removed from the pile without disturbing the rest of the
heap---as in the {\em mikado} game.}) particles. In \fig{fig:1a} peaks are denoted
by gray squares and other particles -- by white ones. The relation of this ballistic
deposition process with the "updating dynamics" on permutations is schematically depicted in
\fig{fig:1b}. We briefly describe it below in Subsection B
and in the Section III in more detail.

\begin{figure}[ht]
\epsfig{file=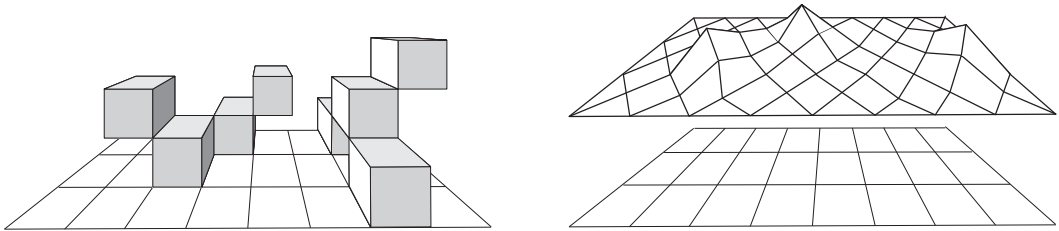,width=14cm} \caption{Two--dimensional ballistic deposition
and corresponding random landscape.} \label{fig:5}
\end{figure}

In a similar fashion, the process of a two--dimensional ballistic deposition in a
box with a square base $L = m \times m$, ($m$ is an integer), can be viewed as a
sequential adding of elementary cubes in the columns satisfying the following rules
(compare to \eq{eq:1}):
\be
h(i,j,n+1)=\max\{h(i-1,j,n),\, h(i+1,j,n),\, h(i,j,n),\, h(i,j-1,n),\,
h(i,j+1,n)\}+1, \label{eq:4}
\ee
where $h(i,j,n)$ is the height of the column with coordinates $(i,j)$ ($1\le
\{i,j\}\le m$) at deposition moment $n$ ($1\le n\le N$). The cubes are added to the
columns with the uniform distribution -- see \fig{fig:5}.

The ``peak'' of a two--dimensional landscape $h(i,j,n)$ is defined
as a local maximum in the set $\{h(i,j,n)\}$ for some fixed moment
$n$, if the following conditions are simultaneously fulfilled:
\be
\left\{\begin{array}{l} h(i,j,n)>h(i-1,j,n) \medskip \\
h(i,j,n)>h(i+1,j,n)
\medskip \\ h(i,j,n)>h(i,j-1,n) \medskip \\ h(i,j,n)>h(i,j+1,n)
\end{array}\right. \label{eq:5}
\ee
Let us note that \eq{eq:4}--\eq{eq:5} imply the periodic boundary conditions both in
$i$ and $j$ coordinates ($1\le \{i,j\}\le m$). The influence of the boundary
condition on the expectation, variance and higher moments of peak numbers can be
easily estimated and becomes negligible in the limit $m \to \infty$.

For these models, our goal is to evaluate the probability distribution function
$P(M,L)$ of having $M$ peaks on a lattice containing $L$ sites. As we proceed to
show, in one--dimension this can be done exactly. Moreover, we are also able to
calculate the ``correlation function'' $p(l)$ defining the conditional probability
that two peaks are separated by the interval $l$ under the condition that the
interval $l$ does not contain peaks. In 2d, determining exactly first three cumulants
of the PDF we define it in the scaling limit using expansion in the Edgeworth series
and show that it converges to a Gaussian function as $L \to \infty$.

\subsection{Interpretation of ballistic growth: "updating dynamics" and random
permutations of natural series}

Let us emphasize that we are interested only in the statistics of peaks of a growing
surface in the \textit{stationary state} and disregard any questions concerning the
statistics of heights.

Dynamics of the set of peaks ${\cal T}$ in the ballistic deposition process depicted
in \fig{fig:1a} can be mapped onto the dynamics of ``peaks'' in the permutation
matrix. We start by describing this connection on an intuitive level. To do this,
let us proceed recursively. Suppose that we deposit a first particle in the column
$i_1$ of an $L$--column box. Next, take the row of $L$ elements with $"1"$ at
position $i_1$ and $"0"$ in all other places: $(\overbrace{0,...,0}^{i_1-1},
1,0,...,0)$. After dropping the second particle, say, in the column $i_2$, take a
row $(\overbrace{0,..., 0}^{i_2-1}, 1,0,...,0)$ and place it over the first one
creating a stack. Suppose that at some time $n$ a new particle is added to the
column which was occupied earlier, say, at time $m$, i.e. $i_n=i_m$ ($n>m$). It
means that we have two identical rows in the stack. In this case, we remove the
first of identical rows (i.e. deposited at time $m$) from the stack and eliminate
the empty line by pulling down all rows deposited after time $m$, as it is depicted
in \fig{fig:1b}d. After some time, the stack will comprise $L$ rows and, according
to the described procedure, will not grow anymore but will be changed only due to
updating of rows (by adding the new ones and by eliminating the old ones). By
construction, this stack is an $L\times L$ permutation matrix. Connecting the
nonzero elements in nearest-neighboring rows by a broken line, we can
straightforwardly define the ``descents'', ``rises'' and ``peaks'' in the
permutation matrix---see \fig{fig:1b}c. The number of peaks at time $N$ in the
$L\times L$ permutation matrix coincides then with the number of peaks in the heap
after having deposited $N$ particles in a box of $L$ columns. The rigorous proof of
this mapping is given in the Section \ref{sect:2:1}.

\begin{figure}[ht]
\epsfig{file=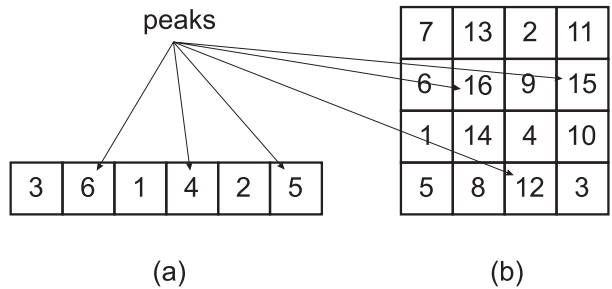, width=8cm} \caption{(a) One--dimensional ($L=6$) and (b)
two-dimensional square ($L=4 \times 4$) lattices with periodic boundary conditions
on sites of which we distribute numbers drawn from the set $1,2,3, \ldots, L$. These
numbers determine the local heights of the surface. The sites with numbers bigger
than those appearing on the neighboring sites are refereed to as the ``peaks''.}
\label{fig:a}
\end{figure}

Consequently, the described relation between the original ballistic deposition model
and an "updating"  dynamics on permutations, depicted in \fig{fig:1b}c,d, allows us
to view the characteristics of the surface obtained within the BD process from a
different perspective: Consider, for example, a one-dimensional lattice containing
$L$ sites, on which we distribute numbers drawn randomly from the set $1,2,3,
\ldots, L$. A number appearing at the site $j$ determines the local ``height''
\footnote{Note that the ``heights'' appearing in permutation matrices have nothing
to do with real heights in the original BD growth model.} of the surface. Now, we
call this site as "a peak", if the number appearing on this site is bigger than the
numbers on two neighboring sites --- see \fig{fig:a}. Generalization to
two-dimensional square lattice with $L = m \times m$, ($m$ is an integer) sites is
straightforward --- see \fig{fig:a}b: the only difference here is that we call as
local ``surface peaks'' such sites $j$ the numbers at which are bigger than numbers
appearing at four adjacent sites. We prove rigorously in what follows that the
probability of having $M$ peaks on a lattice containing $L$ sites on which we place
numbers randomly drawn from the set $1,2,3, \ldots, L$, and  the distribution
function $P(M,L)$ characterizing the number of local surface extrema in the surfaces
of aggregates grown within the BD process, are \textit{identic}. We note finally
that the peak statistics in the latter model with natural numbers $1,2,3, \ldots, L$
and in a related model with a lattice on sites of which one places randomly numbers
uniformly distributed in [0,1] has very interesting and unexpected features, as was
recently communicated to us by B.Derrida \cite{derrida}.

\subsection{Main Results}

Main results of this work are as follows:

\begin{itemize}

\item[(i)] We prove that a uniform one--dimensional ballistic growth process in the
steady state can be mapped onto the "rise-and-descent" sequences in the ensemble of
random permutation matrices.

\item[(ii)] We show that in one dimension the probability distribution function
$P(M,L)$ can be calculated exactly through its generating function
\be
W(s,L) = L! \sum_{M=0}^{\infty} s^{M+1} P(M,L),
\label{mmm}
\ee
where
\be
W(s,L) = \left(\frac{s}{1-\sqrt{1-s}}\right)^{L+1} \sum_{M=0}^{\infty}
\left(\frac{(1 - \sqrt{1-s})^2}{s}\right)^{M+1} A(M,L)
\ee
with $A(M,L)$ being the so-called Eulerian numbers counting the number of ``rises''
(or ``descents'') in the ensemble of equally weighted permutation matrices of size
$L$:
\be
A(M,L)=\sum_{r=0}^M (-1)^r {L+1 \choose r} (M-r)^L, \label{A}
\ee
where ${a \choose b}$ denotes the binomial coefficient; here and henceforth we adopt
a standard notation when ${a \choose b} = \frac{a!}{b!(a-b)!}$ for $b \leq a$ and
${a \choose b} \equiv 0$ otherwise.

Inverting (\ref{mmm}), we find the following exact expression for $P(M,L)$:
\be
P(M,L) = \frac{2^{L+2}}{L!} \sum_{l=1}^M (-1)^{M-l} {\frac{L+1}{2} \choose M-l}
\sum_{m=1}^l \frac{m^{L+1}}{(l-m)! (l+m)!} \label{eq:prob_exact}
\ee
In the limit $L \to \infty$, the PDF $P(M,L)$ converges to a Gaussian distribution:
\be
P(M,L)\sim \frac{3}{2}\sqrt{\frac{5}{\pi L}}
\exp\left\{-\frac{45(M-\frac{1}{3}L)^2}{4L}\right\}.
\label{eq:3b}
\ee
Note that a similar result has been previously found in Ref.\cite{ov} for the sum of
peaks and throughs in a permutation of length $L$ within the context of the number
of the U-turns made by the "Permutation Generated Random Walk" (PGRW).

\item[(iii)] We show that in one--dimensional systems the probability $p(l)$ of having
two peaks  separated by the interval $l$ under the condition that this interval does
not contain other peaks, obeys:
$$
p(l)=2^{l}\frac{(l-1)(l+2)}{(l+3)!}
$$
Curiously enough, this expression coincides with the distribution function of the
distance between two "weak" bonds obtained by Derrida and Gardner \cite{derrida2} in
their analysis of the number of metastable states in a one--dimensional random Ising
chain at zero's temperature.

\item[(iv)] Using the cumulant expansion, we show that in two dimensions, in the limit
$L \to \infty$, the normalized probability distribution function $p(x, L)$, where $x
= (M - \mu_1^{2D})/\sigma$, converges to a Gaussian distribution
$$
p\left(x = \frac{M - \mu_1^{2D}}{\sigma}, L\right) \to \frac{1}{\sqrt{2 \pi}}
e^{-x^2/2} \left(1+\frac{1}{\sqrt{L}}f(x)+ o\left(\frac{1}{\sqrt{L}}\right)\right),
$$
where
$$
f(x)=\sqrt{L} \frac{\kappa_{3}^{\rm 2D}} {\left(\kappa_{2}^{\rm 2D} \right)^{3/2}}
\frac{1}{6} (x^{3}-3 x)
$$
and first three cumulants are given by
$$
\left\{\barr{l} \disp \kappa_{1}^{\rm 2D}=\mu_{1}^{\rm 2D}=\frac{1}{5}L \medskip
\\ \disp \kappa_{2}^{\rm 2D}=\mu_{2}^{\rm 2D}=\sigma^{2}= \frac{13}{225}L \medskip \\
\disp \kappa_{3}^{\rm 2D}=\mu_{3}^{\rm 2D}=\frac{512}{32175} L \earr \right.
$$

\item[(v)] We have realized that local peaks in one-- and  two--dimensional
equilibrium NNN ballistic deposition model are correlated and the correlation length
extends over two lattice spacings: while by definition, the probability of having
two peaks at the neighboring sites is zero, appearance of two peaks on
next-nearest-neighboring sites (also on a diagonal for 2d) is higher than squared
mean density of peaks ($1/9$ and $1/25$ in 1d and 2d, respectively). On contrary,
the probability of having two peaks at a distance exceeding two lattice spacings is
equal to the squared mean density, which signifies that they are uncorrelated. In
other words, there are effective short-range "attractive" interactions between
peaks. The corresponding weights of a few typical configurations are summarized in
figure \fig{fig:prob}.

\begin{figure}[ht]
\epsfig{file=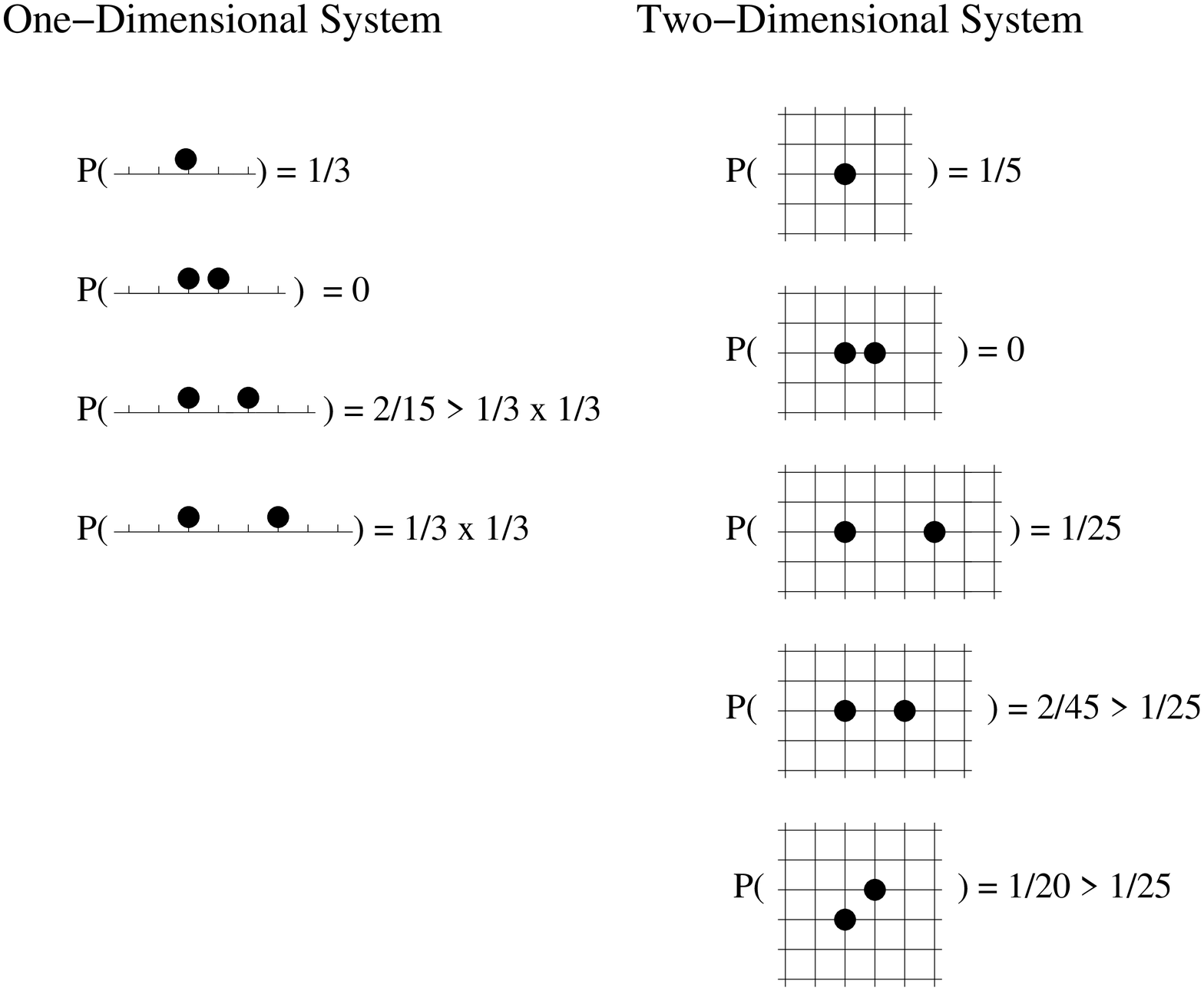, width=12cm} \caption{Probabilities of having a single
peak and a pair of isolated peaks some distance apart from each other in one-- and
two--dimensional models. Filled circles denote peaks - sites with numbers exceeding
numbers appearing on nearest-neighboring sites. First column presents probabilities
of several configurations in 1d, while the second one describes the probabilities of
a few possible configurations appearing in two dimensions.} \label{fig:prob}
\end{figure}

\end{itemize}

\section{Rigorous description of the relation between ``updating dynamics'' on
permutations and BD.}
\label{sect:2:1}

As we have already mentioned in the previous section, dynamics of the set of peaks
${\cal T}$ in the ballistic deposition process depicted in \fig{fig:1b} can be
mapped onto the dynamics of ``peaks'' in the permutation matrix. Let us formulate
this mapping explicitly.

In a more rigorous approach, we have to consider two dynamical systems: one -- on
peak sets and the other -- on permutations. Let us call $E_i$ the operation which
corresponds to dropping a box in the column $i$ on a peak set. That is if $S$ is a
peak set (i.e. a subset of $\{1,\dots,L\}$ without two consecutive numbers) then
\be
E_i(S) := S \cup \{i\} / \{i-1, i+1\}\,. \ee The corresponding operation $F_i$ on
permutations is defined by \be \mu := F_i(\sigma)\qquad\text{with}\qquad \mu(k) :=
\begin{cases}
\sigma(k) & \text{if $\sigma(k) < \sigma(i)$} \\
L & \text{if $k = i$} \\
\sigma(k)-1 & \text{if $\sigma(k) > \sigma(i)$} \\
\end{cases}
\ee
To proceed, we also need to define the reverse operation on permutations, $R_k$,
which removes the upmost row and inserts it under the row $k$:
\be
\mu := R_k(\sigma)\qquad\text{with} \qquad \mu(l) :=
\begin{cases}
\sigma(l) & \text{if $\sigma(l) < k$} \\
k & \text{if $\sigma(l) = L$} \\
\sigma(l)+1 & \text{if $L > \sigma(l) > k$} \\
\end{cases}
\ee
It is clear now that $F_i(\sigma) = \mu$ if and only if $R_k(\mu) = \sigma$ where
$k=\sigma(i)$ is the row of the $1$ in the column $i$.

Denote now by $\Peak(\sigma)$ the peak set of the permutation $\sigma$. Then, it is
obvious that for all permutation $\sigma$
\be
E_i(\Peak(\sigma)) = \Peak(F_i(\sigma))\qquad\text{for all $i\leq L$}.
\label{equ:EiFiPeak}
\ee
This simple but crucial observation allows us to translate the dynamics on
permutations to the dynamics of peak sets. We will proceed in two steps: first, we
will show that the limit probability measure of the dynamical system on permutations
is equi-distributed and second, we will demonstrate that the image of this
probability measure by the peak--set map is the limit probability measure on peak
sets.

Let $(P(\sigma))_\sigma$ be a probability measure on permutations. Then, after one
uniformly random chosen step $F_i$ the new probability $P'(\mu)$ of a permutation
$\mu$ is
\be
P'(\mu) = \frac1L \sum_{i=1}^L \sum_{\sigma} P(\sigma)
\ee
where the inner sum extends over the sets of all permutations $\sigma$, such
that $F_i(\sigma) = \mu$. However, by definition of $F_i$, if a permutation
$\tau$ is of the form $\tau = F_i(\sigma)$ then $\tau(i) = L$. Thus in the
previous sum the only value of $i$ such that there exists some $\sigma$ such
that $F_i(\sigma) = \mu$ is $i = \mu^{-1}(L)$, that is if $i$ is the column
where the $1$ is in the top row $L$ in $\mu$. Now, here are exactly $L$
permutations $\sigma$ verifying $F_i(\sigma) = \mu$, namely $R_k(\mu)$ for
$k=1\dots L$. Consequently, $P'(\mu)$ is simply the average of the
probabilities of the $R_k(\mu)$:
\be
P'(\mu) = \frac1L \sum_{k=1}^L P(R_k(\mu)). \label{equ:PprimeP}
\ee
Moreover, it is clear that for all pairs of permutations $\sigma,\mu$ there is a
sequence $(i_r)_r$ of transformations which maps $\sigma$ on $\mu$, that is $\mu =
\dots F_{i_3} F_{i_2} F_{i_1}(\sigma)$.  Consequently, the map $F:P\mapsto P'$ is an
irreducible Perron--Frobenius map whose maximal module eigenvalue is $1$ with
multiplicity $1$. Thus, when the number of boxes tends to infinity, the probability
converges to some limit and this limit is the unique, normalized (the sum of the
coordinate is $1$) eigenvector of $F$ corresponding to the eigenvalue $1$. In virtue
of \eq{equ:PprimeP}, the uniform distribution on permutations is preserved by $F$ so
that it must be the limit distribution.

Using \eq{equ:EiFiPeak}, it is possible to translate our permutations language to
the language of the peak sets. Recall that we call a peak set any subset of
$\{1,\dots,L\}$ without two consecutive numbers. The only needed remark is that any
non empty peak set $S$ is the peak set of a permutation. Thus the map $\Peak$
extends to a surjective map from the set of probability measure on permutations to
the set of probability measures on peak sets:
\be
\Peak : P \longmapsto \Peak(P)(S) := \sum_{\sigma\ :\ \Peak(\sigma) = S}
P(\sigma)\,.
\ee
Denote next $E$ the map $P(S)\mapsto P'(S)$ where $P'(S)$ is the new probability
measure on sets after a uniformly random step $E_i$. Then \eq{equ:EiFiPeak} can be
written down as
\be
E(\Peak(P)) = \Peak(F(P)) \,, \label{equ:EFPeak}
\ee
for all probability measures on permutations $P$. In particular, the spectrum of $E$
is included in the spectrum of $F$ and moreover, the generalized eingenspaces
(kernel of $(F-\lambda I)^m$ for $m$ large) of $E$ are the image under $\Peak$ of
the generalized eigenspaces of $F$. Consequently, $E$ is a Perron--Frobenius map
with maximal module eigenvalue $1$ with multiplicity $1$. As a consequence the limit
distribution on peak sets exists and is the image by the map $\Peak$ of the limit
distribution on permutations
\be
P_{\text{limit}}(S) = \Peak\left(P_{\text{limit}}(\sigma)\right)(S) = \frac1{L!}
\card\{ \sigma\ |\ \Peak(\sigma) = S \}\,. \label{equ:peakFormula}
\ee

Before we proceed further, a few comments might be in order:

First, as a matter of fact, this technique is quite general and can be applied to a
more general notion of heaps of pieces. We demonstrate it on the following example.
Let $G = (V,E)$ be a finite oriented simple graph with vertex and edge set $V$ and
$E$ and let $L$ be the number of vertices. Each vertex $v\in V$ is associated with
one "type" of pieces, and an arrow $e = v \mapsto v'$ encodes the fact that when a
piece of type $v$ falls after a piece of type $v'$, then it is placed over it, thus
they are no more pieces of type $v'$ at the top of the pile. The notion of
permutation generalizes to the notion of standard labeling of the vertex of the
graph, that is assignation of distinct numbers from $1\dots L$ to the vertices $v$.
Then a vertex $v$ is called a peak of a labeling $l$ if $l(v) > l(v')$ for all edges
$v \mapsto v'$. One immediately notices that the previous reasoning applies, which
implies that for all sets $S$ of vertices the limit probability measure
$P_{\text{limit}}(S)$ of $S$ to be exactly the set of maximal pieces of a random
heaps on the graph $G$ is given by \eq{equ:peakFormula}. In this regard, the limit
distribution of the 2D models can also be computed using these generalized peaks.
Moreover, in the case of the oriented lines the notion of peaks reduces to the
notion of a descents.

Second, combinatorics of peaks of permutations has been recently reviewed by J.
Stembridge in Ref.\cite{stem}. The peak algebra of Stembridge is a certain
sub-algebra of the algebra of polynomials with infinitely many variables. It has a
natural basis $K_S$ indexed by peak sets $S$ and therefore can serve as the support
for generating series of probability of a peak set. In this regard, the generating
series of the probability of each peak set has a very simple expression
\be
\sum_{L=0}^\infty t^L \sum_{S \subset \{1\dots L\}} P(S) K_S  = \sum_{L=0}^\infty
\frac{t^L}{L!} \sum_{S \subset \{1\dots L\}} \card\{ \sigma\ |\ \Peak(\sigma) = S
\}\, K_S = \exp (K_1 t)
\ee
where $K_1$ is the unique peak set where $L=1$. Note that there is a minor
difference between our work and Ref.\cite{stem} since in the latter the extremities
$1$ and $L$ were never considered as a peak, while we use the periodic boundary
conditions.  This difference should be unimportant for sufficiently large $L$.

Furthermore it should be noticed that in \cite{billHVW} a different random walk is
considered on peaks and conjectured on permutations with the same limit probability
measure. Finally, the result of \cite{BHT} suggests that this measure should be seen
as some kind of generalized Plancherel measure associated with the degenerated
Hecke--Clifford algebra instead of the symmetric group.

\section{Probability distribution function $P(M,L)$ in 1d}

We are now in position to determine exactly the probability $P(M,L)$ that a random
heap has a fixed number of peaks. From our previous analysis, it follows that such a
probability is equal to the number $B(M,L)$ of permutations  of length $L$ having
exactly $M$ peaks, divided by the total number of permutations $L!$, i.e.
$P(M,L)=B(M,L)/L!$. Now, we recollect that if one considers permutation descents
instead of peaks, the numbers $A(M,L)$ of permutations of $1\dots L$ with exactly
$M$ descents are known as the Eulerian numbers, obey the following three--site
recursion
\be
A(M,L) = (L-M+1)\, A(M-1,L-1) + M\, A(M,L-1)\, \label{eq.euler}
\ee
which leads to the following recurrence relation for the
generating function $U(t,L)= \sum_{M=0}^{\infty} t^{M+1} A(M,L)$:
\be
U(t,1) = t, \qquad U(t,L) = t(1-t)\frac{d}{dt} U(t,L-1) + L t
U(t,L-1)\,.
\ee

The recursion relation for peak--Eulerian number $B(M,L)$ has been
considered by Stembridge in \cite{stem}, Remark 4.8. In our
notations, his results attains the following form
\be
B(M,L) = (L - 2M + 2)\, B(M-1,L-1) + 2 M\, B(M,L-1)\, \label{eq.eulerPeak}
\ee
which in turn can be encoded by the recurrence relation obeyed by the generating
series $W(s,L) =\sum_{M=0}^{\infty} s^{M+1} B(M,L)$ (note that compared to
Stembridge there is no $t^{k+1}$ term):
\be
W(s,1) = s, \qquad W(s,L) = 2s(1-s) \frac{d}{ds} W(s,L-1) + L s W(s,L-1)\,.
\label{stem}
\ee
Here is a table of a few first polynomials $W(s,L)$:
\be
\begin{array}{lcl}
W(s,1) & = & s \medskip  \\
W(s,2) & = & 2\, s \medskip \\
W(s,3) & = & 2\, s^2 + 4\, s \medskip \\
W(s,4) & = & 16\, s^2 + 8\, s \medskip \\
W(s,5) & = & 16\, s^3 + 88\, s^2 + 16\, s \medskip \\
W(s,6) & = & 272\, s^3 + 416\, s^2 + 32\, s \medskip \\
W(s,7) & = & 272\, s^4 + 2880\, s^3 + 1824\, s^2 + 64\, s \medskip \\
W(s,8) & = & 7936\, s^4 + 24576\, s^3 + 7680\, s^2 + 128\, s \medskip \\
W(s,9) & = & 7936\, s^5 + 137216\, s^4 + 185856\, s^3 + 31616\, s^2 + 256\, s
\end{array}
\label{gen_f}
\ee
Differentiating polynomials $W(s,L)$ and setting $s = 1$, we can find, in principle,
any moments of the probability distribution function. For example, the expectation
of the number of peaks is $\frac{L+1}{3}$ for $L \geq 2$; the variance is
$\frac{2L+2}{45}$ for $L \geq 4$, while the third central moment is equal to
$-\frac{2L+2}{945}$ for $L \geq 8$.

\subsection{Permutation Generated Random Walks}
\label{sect:2:2}

In this subsection we briefly outline the basic notions concerning the so-called
Permutation Generated Random Walk (PGRW) proposed in Ref.\cite{ov}. Consider a given
permutation $\pi=\{\pi_1,\pi_2,\pi_3,...,\pi_{L+1}\}$ of $[L+1]=1,2,3, \ldots, L+1$
and rewrite it as a 2--line table:
$$
\pi=\left(\begin{array}{ccccc} 1 & 2 & 3 & ... & L+1 \\ \pi_1 & \pi_2 & \pi_3 & ...
& \pi_{L+1} \end{array} \right).
$$
Suppose that this table assigns some discrete ``time'' variable $j$ ($j=1,2,.3,...,
L+1$, upper line in the table) to each permutation encountered in the second line
and, hence, allows to order this permutation.

Now, in a standard notation, we call $j$ the ``rise'', if $\pi_j<\pi_{j+1}$,
otherwise, if $\pi_j>\pi_{j+1}$, we refer to it as the ``descent''. Further on, if
we have simultaneously $\pi_{j-1}<\pi_j$ and $\pi_j>\pi_{j-1}$, we call $j$ - the
``peak''.

Then, the Permutation Generated Random Walk is defined by the following recursive
procedure:
\begin{enumerate}
\item[(i)] at time moment $j=0$ the walker is at the origin;
\item[(ii)] at time $j > 0$ the walker makes a step to the right if $j$ is the rise,
and makes a step to the left if $j$ is the descent.
\end{enumerate}
Note that, evidently, if $j$ is a peak, the walker makes a left``U-turn''.

Statistical properties of the PGRW were studied in Ref.\cite{ov},
where the distribution of the end--to--end distance and intermediate
points of the trajectory, probability measure of different
trajectories, the number of U-turns, as well as various correlation
functions have been analyzed with respect to the uniform measure on
the ensemble of random permutations.

Using the methods developed in Ref.\cite{ov}, and exploiting the connection between
the one--dimensional ballistic deposition process and dynamics on permutations
established in the previous section, one can straightforwardly reconstruct the
probability distribution $P(M,L)$ of having $M$ peaks in the uniform ballistic
deposition process in the planar box composed of $L$ columns in the stationary
regime. According to \cite{stem}, the generating function of number of peaks
$W(s,L)$ can be expressed in terms of the generating function of the number of
rises, $U(t,L)$, in the ensemble of equally weighted permutation matrices of size
$L$ (the so-called Eulerian number), Eq.(\ref{A}). The explicit relation between
$W(s,L)$ and $U(t,L)$ is:
\be
W(s,L) = \left(\frac{2}{1+t}\right)^{L+1} U(t,L)
\label{eq:gen_fn} \ee where
\be s=\frac{4t}{(1+t)^2}\,; \qquad  t = \frac{2-s-2\sqrt{1-s}}{s}=
\frac{(1-\sqrt{1-s})^2}{s} \label{eq:st}
\ee

Taking into account the identity established in \cite{ov}:
\be \label{a}
A(M,L) = \frac{2L!}{\pi} \int_0^{\infty} \left(\frac{\sin
x}{x} \right)^{L+1} \cos(x(2M-L-1))\,,
\ee
we arrive at the following explicit expression for the generating function $W(s,L)$:
\be
\begin{array}{lll}
W(s,L) & = & \disp \left(\frac{2}{1+t}\right)^{L+1} \sum_{M=1}^{\infty} t^{M+1}
A(M,L) \medskip \\ & = & \disp \frac{2^{L+1}L!}{\pi (1+t)^{L+1}}\,
\int\limits_{0}^{\infty} dx \left(\frac{\sin x}{x} \right)^{L+1}
\frac{e^{-ix(L+1)}t\left((1+e^{2ix(3+L)})t-e^{2ix(2+L)}-
e^{2ix}\right)}{(1+e^{4ix})t-e^{2ix}(1+t^2)}
\label{eq:w}
\end{array}
\ee
with $t$ given by \eq{eq:st}. Polynomials $W(s,L)$ admit also another useful
representation directly following from the relations between generating functions
$W(s,L)$ and $U(t,L)$ found in \cite{stem}:
\be
W(s,L) = 2^{L+2} (1-s)^{(L+1)/2}\, \mbox{Li}_{-(L+1)}
\left(\frac{(1-\sqrt{1-s})^2}{s}\right) \label{eq:polylog}
\ee
where
$\mbox{Li}_{\nu}(x)$ is the Poly-log function.

\begin{figure}[ht]
\epsfig{file=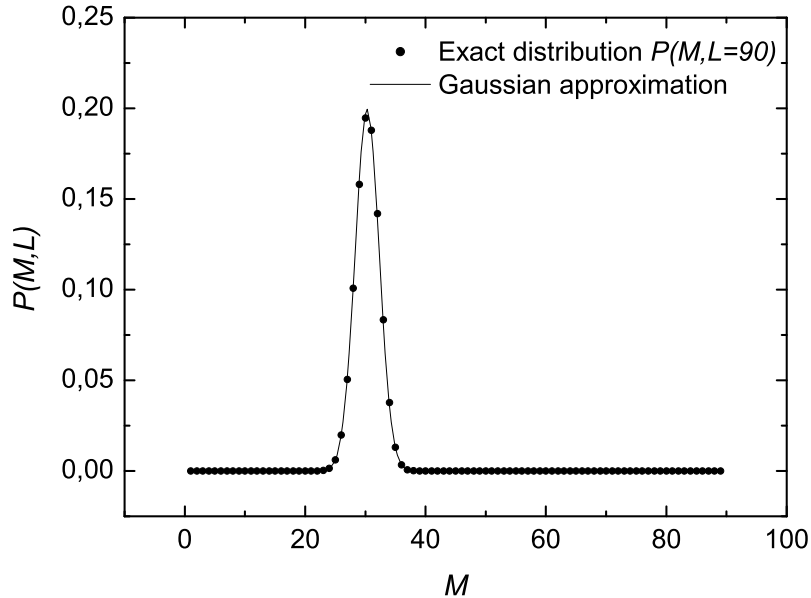, width=12cm} \caption{Exact distribution function
\eq{eq:prob_exact} ($\bullet$) and its Gaussian approximation \eq{eq:3b} (line) for
$L=90$.} \label{fig:peaks}
\end{figure}

Exact expression for the distribution function $P(M,L)$ obtained from
\eq{eq:polylog} is given by Eq.(\ref{eq:prob_exact}). More details on its derivation
are presented  in the Appendix. In the asymptotic limit $L\gg 1$ the double sum
\eq{eq:prob_exact} converges to a Gaussian function with a nonzero mean \eq{eq:3b}
--- see Fig.\ref{fig:peaks}.

Calculating the  first three central moments of the limiting
distribution function in \eq{eq:3b} for $L\gg 1$, we get:
\be
\left\{ \barr{ccc} \disp \mu_{1}^{\rm 1D} =
\sum_{M=1}^L M\, P(M,L)  &= & \disp \frac{1}{3}L, \medskip \\
\mu_{2}^{\rm 1D}= \left<\Big(M - \la M\ra \Big)^2\ra & = & \disp \frac{2}{45}L,
\medskip \\ \mu_{3}^{\rm 1D}= \left<\Big(M - \la M\ra \Big)^3\ra & = & 0. \earr
\label{eq:3c} \right.
\ee

Note that the first two expressions coincide with the expectation and the variance
computed by J.Desbois in \cite{des} using decoupling of the hierarchy of correlation
functions. The third central moment, i.e. $\mu_3^{\rm 1D}$, of the distribution
$P(M,L)$, naturally turns to be equal to 0 since Gaussian limit is considered. In
what follows we proceed to show using the method of the correlation functions (see
the next Section for details) that all moments of the distribution may be calculated
exactly, for arbitrary finite $L$. As a matter of fact, the value of $\mu_3^{\rm
1D}$ is
\be
\mu_3^{\rm 1D}=\la \Big(M - \la M\ra \Big)^3\ra = -\frac{2}{945}L, \label{eq:3d}
\ee i.e. it is not zero and grows
linearly with the system size $L$. This value of $\mu_{3}^{\rm 1D}$ coincides with
the one computed using the generating function in \eq{gen_f} or \eq{eq:w}.

\subsection{Correlation functions of the PGRW within the operator formalism}
\label{sect:2:3}

In  Ref.\cite{ov} it has been shown that a non-Markovian PGRW can be effectively
described using an auxiliary, recursively constructed Markovian process, which has
the same distribution as the PGRW. Employing the ideas of Hammersley \cite{ham} (see
also \cite{ald} for more details), who has used such an approach in his celebrated
analysis of the problem of the longest increasing subsequence in random
permutations, the following random walk model has been constructed:
\begin{itemize}
\item Suppose one has an infinite in both directions line of integers and a random
walker initially at the origin;
\item Trajectory $Y_l$ of this auxiliary process is built step by step: at time
moment $l$ define a real-valued random variable $x_{l+1}$, having a uniform
distribution in $[0,1]$. If $x_{l+1} > x_{l}$ the walker is moved to the right, i.e.
$Y_l = Y_{l-1} + 1$; otherwise, it goes to the left, i.e. $Y_l = Y_{l-1} - 1$. At
the next time moment, chose $x_{l+2}$, compare it with $x_{l+1}$ and move the walker
accordingly, and etc.
\end{itemize}

Two strong results have been proven in Ref.\cite{ov} concerning the relation between
this recursively constructed Markovian process and the PGRW: First of all, it has
been shown that the probability $P(Y_L = X)$ for the trajectory $Y_L$ of the
auxiliary process to appear at site $X$ at time moment $L$ is equal to the
probability $P(X_L^{(L)}= X)$ for the end--point $X_L^{(L)}$ of the PGRW trajectory
$X_l^{(L)}$ generated by a given permutation of $[L+1]$ to appear at site $X$. This
signifies that
\be
P(Y_L = X) = \frac{(1-(-1)^{X+L})}{2 L!} A\left(\frac{X+L-1}{2}, L\right).
\ee
(Recall that $A(...)$ is the Eulerian number).

Second, it has been proven that the probability $P(X_l^{(L)} = X)$ for the PGRW
trajectory $X_l^{(L)}$ at intermediate time moment $l = 1,2,3, \ldots,L$ to appear
at site $X$ is equal to the probability $P(Y_l = X)$ for the trajectory of the
auxiliary process $Y_l$ to appear at time moment $l$ at site $X$. That is,
\be
P(X_l^{(L)} = X) = P(Y_l = X) = P(X_L^{(L)} = X) = \frac{(1-(-1)^{X+l})}{2 l!}
A\left(\frac{X+l-1}{2},l\right),
\ee
which signifies that distribution of any intermediate point $X_l^{(L)}$ of the PGRW
trajectory generated by permutations of a sequence of length $L+1$ depends on $l$
but is independent of $L$.

Using this equivalence, it is thus possible to write down explicitly the probability
measure of any given PGRW trajectory (or of some part of it) as a chain of iterated
integrals over real-valued random variables $x_{l}$. Recollecting the definition of
the PGRW and noticing that each trajectory mirrors one-by-one a unique
"rise-and-descent" sequence in a random permutation, we are thus able to represent a
probability of any "rise-and-descent" sequence as a chain of iterated integrals over
real-valued random variables $x_{l}$ uniformly distributed in $[0,1]$.

More specifically, following Ref.\cite{ov}, consider a given "rise-and-descent"
sequence $\alpha(L)$ of length $L$ of the form:
$$
\alpha(L)=\left(\begin{array}{ccccc} 1 & 2 & 3 & ... & L \\
a_1 & a_2 & a_3 & ... & a_{L}
\end{array} \right).
$$
where $a_l$ can take either of two symbolic values: $\up$ or $\down$. Consequently,
the first line in the table is the running index $l$ which indicates position along
the permutation, while the second line shows what we have at this position - a rise
or a descent. Assign next to each symbol at position $l$ an integral operator; for a
rise ($\up$) it is $I_l(\up)$, while for a descent ($\down$) it will be
$I_l(\down)$, where \be \hat{I}_l(\up)= \int_{x_{l-1}}^1 dx_{l} \qquad \mbox{and}
\qquad \hat{I}_l(\down)=\int_0^{x_{l-1}} dx_{l}. \label{eq:cor1} \ee To each
$L$--step trajectory one associates next a characteristic polynomial $Q(x,
\alpha(L))$ defined as an "ordered'' product \cite{ov}: \be Q(x,\alpha(L))=\; :
\hspace{-4pt} \prod_{l=1}^L \hspace{-4pt} : \; \hat{I}_l(a_l) \cdot 1
\label{eq:cor2}, \ee where $a_l=\{\up,\down\}$ for $l=1,...,L$. The statistical
weight, i.e. the probability distribution function, $P(\alpha(L))$, of this given
"rise--and--descent" sequence $\alpha(L)$ in the ensemble of all equally likely
permutations is then simply given by \be P(\alpha(L))=\int_0^1 Q(x,\alpha(L)) dx .
\label{eq:cor2a} \ee

\begin{figure}[ht]
\epsfig{file=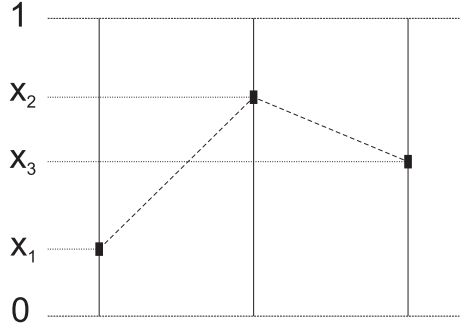,width=6cm} \caption{Three markers creating a peak
configuration.} \label{fig:2}
\end{figure}

In may be expedient to illustrate this formal representation on a
particular example. Consider, e.g., a given "rise-and-descent"
sequence of the form $\{\up,\up,\down,\up,\up\}$. For this sequence,
the characteristic polynomial $Q(x,\alpha(5))$ reads:
$$
Q(x,\alpha(5))= \hat{I}_1(\up)\, \hat{I}_2(\up)\,\hat{I}_3(\down)\, \hat{I}_4(\up)\,
\hat{I}_5(\up) \cdot 1 = \int\limits_x^1 dx_1 \int\limits_{x_{1}}^{1} dx_2
\int\limits_{0}^{x_{2}} dx_3 \int\limits_{x_3}^1 dx_4 \int\limits_{x_{4}}^{1} dx_5
\cdot 1= \frac{3}{40} -\frac{x}{8}+\frac{x^{3}}{12}-
\frac{x^{4}}{24}+\frac{x^{5}}{120},
$$
and, hence, the probability of this particular configuration is
$$
P(\alpha(5))=\int_0^1 Q(x,\alpha(5))\,dx =\frac{19}{720} .
$$
We note that construction of the probability measure in
\eq{eq:cor1}--\eq{eq:cor2a} can be most easily understood by
considering the following example. Suppose there are three
``markers'' representing the particles with the coordinates $x_1,\,
x_2,\, x_3$ ($0\le \{x_1,\, x_2,\, x_3\}\le 1$)---see \fig{fig:2}.
Markers in \fig{fig:2} can be independently deposited in the
interval $[0,1]$ with uniform distribution. It is obvious that the
probability $P(\up\,\down)$ for three particles to create a peak, is
defined by the probability of a configuration with $0\le x_1<x_2$
and $x_2>x_3\le 1$. Thus,
$$
P(\up\,\down)=\int_0^1 dx_3 \int_{x_3}^1 dx_2 \int_0^{x_2} dx_1
\cdot 1 = \frac{1}{3},
$$
what coincides with the operator expression
$$
P(\up\,\down) = \int_0^1 \hat{I}_1(\up)\, \hat{I}_2(\down)\,dx.
$$
(compare to \eq{eq:cor1}--\eq{eq:cor2}).

\subsection{Exact calculation of the first three central moments of the distribution $P(M,L)$ in 1d.}
\label{sect:2:4}

Now we apply the operator formalism for the computation of three
first central moments of the probability distribution function,
$P(M,L)$, of having $M$ peaks on a one--dimensional periodic lattice
of size $L$: \be \barr{l}
\disp \mu_1^{\rm 1D} \equiv \la M \ra \medskip, \\
\mu_2^{\rm 1D} \equiv \la M^2 \ra - \la M \ra^2 \medskip, \\
\mu_3^{\rm 1D} \equiv \la M^3 \ra - 3 \la M \ra \la M^2 \ra + 2 \la M \ra^3. \earr
\label{eq:cor3}
\ee

Let us introduce
\be
\Delta_{i}^{-}=\theta(x_{i}-x_{i-1}) \qquad \mbox{and} \qquad \Delta_{i}^{+}=
\theta(x_{i}-x_{i+1}) \label{eq:Delta},
\ee
for $i=1,...,L$, where $\theta(x)$ is the usual Heaviside step--function
$$
\theta(x) = \left\{\barr{ll} 1 & \quad \mbox{for $x>0$}, \medskip \\
 0 & \quad \mbox{for $x<0$}. \earr \right.
$$

Because of the periodic boundary conditions we have
$\Delta_{1}^{-}=\theta(x_{1}-x_{L})$ and $\Delta_{L}^{+}=\theta(x_{L}-x_{1})$. Using
the operator formalism we can represent the expectation $\mu_1^{\rm 1D}$ in the
following form \be \mu_{1}^{\rm 1D} = \int \limits_{0}^{1} \int \limits_{0}^{1} ...
\int \limits_{0}^{1} \left[\sum \limits_{i=1}^{L} \Delta_{i}^{-}\Delta_{i}^{+}
\right] dx_{1}dx_{2}... dx_{L}. \label{eq:mu1d1} \ee In the sum above each term
$\Delta_{i}^{-} \Delta_{i}^{+}= \theta(x_{i}-x_{i-1}) \theta(x_{i}-x_{i+1})$
corresponds to the peak at the position $x_{i}$. All $L$ terms in the sum in
\eq{eq:mu1d1} are identical and independent, so it is possible to rewrite
\eq{eq:mu1d1} as \be \mu_{1}^{\rm 1D} = L \int \limits_{0}^{1} \int \limits_{0}^{1}
\int \limits_{0}^{1} \theta(x_{i}-x_{i-1})\theta(x_{i}-x_{i+1})
dx_{i-1}dx_{i}dx_{i+1} =L \int \limits_{0}^{1} \left[ \int \limits_{0}^{x_{i}} \int
\limits_{0}^{x_{i}} dx_{i-1}dx_{i+1} \right] dx_{i}=\frac{1}{3}L . \label{eq:mu1d2}
\ee

\begin{figure}[ht]
\epsfig{file=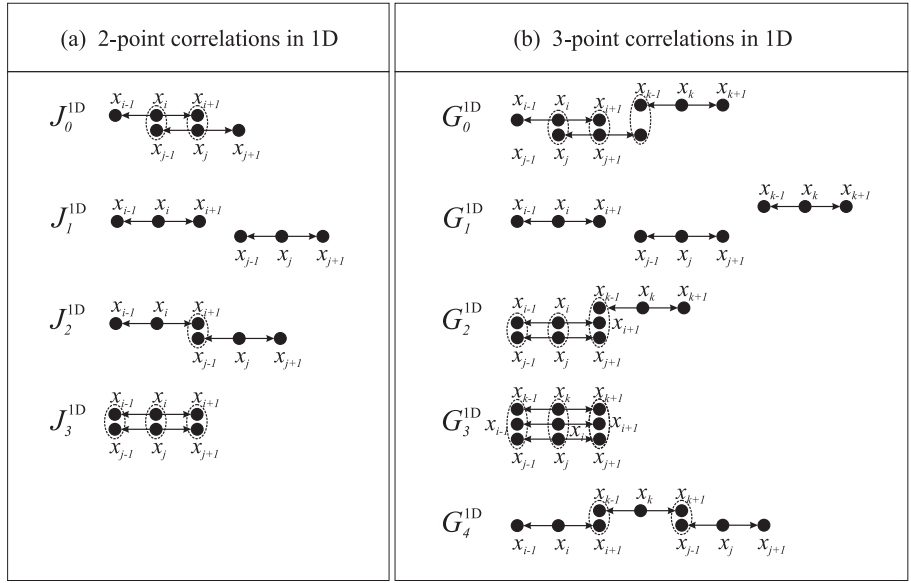,width=12cm} \caption{Basic diagrams and corresponding
integrals: a) $J_{r}^{1D}$ for the computation of $\mu_{2}^{\rm 1D}$; b)
$G_{r}^{1D}$ for the computation of $\mu_{3}^{\rm 1D}$.} \label{fig:fig5}
\end{figure}

The second central moment $\mu_2^{\rm 1D}$ can be calculated as
follows \be \barr{lll} \disp \mu_2^{\rm 1D} & \equiv & \disp \la M^2
\ra - \la M \ra^2 = \int \limits_{0}^{1}... \int\limits_{0}^{1}
\left[\sum \limits_{i=1}^{L} \sum \limits_{j=1}^{L} \Delta_{i}^{-}
\Delta_{i}^{+} \Delta_{j}^{-}\Delta_{j}^{+}\left\{
\delta(x_{a}-x_{b}) \right\}\right]
dx_{1}dx_{2}... dx_{L}- \frac{1}{9}L^{2} = \medskip  \\
& = & \disp L\sum \limits_{j=1}^{L}\left[\int\limits_{0}^{1}... \int \limits_{0}^{1}
\Delta_{i}^{-}\Delta_{i}^{+} \Delta_{j}^{-} \Delta_{j}^{+}
\left\{\delta(x_{a}-x_{b}) \right\} dx_{i-1}dx_{i}dx_{i+1} dx_{j-1}dx_{j}dx_{j+1} -
\frac{1}{9}L\right]
\medskip \\ & = & \disp  L \left[\sum \limits_{r=0}^{3} a_{r}^{\rm 1D}J_{r}^{\rm 1D}
-\frac{1}{9}L \right], \earr \label{eq:mu2d1}
\ee
where the notation $\{\delta(x_{a}-x_{b})\}$ means the following: if some point
$x_a\in \{x_{i-1},x_i,x_{i+1}\}$ coincides with another point $x_b\in
\{x_{j-1},x_j,x_{j+1}\}$, then $\{\delta(x_{a}-x_{b})\}=1$ otherwise
$\{\delta(x_{a}-x_{b})\}=0$ and we sum over all configurations of diagrams shown in
\fig{fig:fig5}a. For example, if in some configuration the points $x_{i+1}$ and
$x_{j}$ coincide, then $\{\delta(x_{a}-x_{b})\}\equiv \delta(x_{i+1}-x_{j})$ etc.
The value $J_{r}^{\rm 1D}$ is the integral of the diagram $r$, $a_{r}^{\rm 1D}$ is
the "weight" of the corresponding diagram (i.e. the number of identical diagrams).
The graphic representation of integrals in eq.\eq{eq:mu2d1} is given in
\fig{fig:fig5}a. We consider the system of length $L$ with periodic boundary
conditions. Instead of summing over $L$ possible values of $i$ we fix some arbitrary
value $x_{i}$ and perform averaging over all possible positions of $x_{j}$. That
gives us $L$ in front of the sum in \eq{eq:mu2d1}. The integral in \eq{eq:mu2d1}
depends on $j-i$ only. We enumerate all possible values of the integral by the index
$r$, and compute the weight $a_{r}^{\rm 1D}$ of each integral $J_r^{\rm 1D}$ in the
sum for $j=1,2,..., L$.

The total number of diagrams is normalized: $\sum \limits_{r=0}^{3} a_{r}^{1D}=L$.
The computed values of $J_r^{\rm 1D}$ and of $a_r^{\rm 1D}$ are collected in the
Table \ref{tab:2crosd}. All such configurations are generated by shifting the point
$x_j$ with respect to $x_i$ as shown in Fig.\ref{fig:fig5}a,b.

\begin{table}[ht]
\begin{tabular}{|l|c|c|c|c|} \hline $r$ & 0 &
1 & 2 & 3 \\ \hline $J_r^{\rm 1D}$ & 0 & $\frac{1}{9}$ & $\frac{2}{15}$ &
$\frac{1}{3}$
\\ \hline $a_{r}^{\rm 1D}$ & 2 & $L-5$ & 2 & 1 \\ \hline
\end{tabular}
\caption{Values of diagrams $J_r^{\rm 1D}$ and of the corresponding weights
$a_{r}^{\rm 1D}$ for computing $\mu_{2}^{\rm 1D}$.} \label{tab:2crosd}
\end{table}

If the shortest distance between $x_{i}$ and $x_{j}$ is larger than
2 (respecting the periodic boundary conditions), then the integrals
over $\Delta_{i}^{-}\Delta_{i}^{+}$ and
$\Delta_{j}^{-}\Delta_{j}^{+}$ decouple and give $J_{1}^{\rm
1D}=[P(\up\down)]^{2}= \frac{1}{3}\times\frac{1}{3} =\frac{1}{9}$.
The contribution of these integrals cancels by subtracting
$\frac{1}{9}$. The number of such integrals is $a_{1}^{\rm 1D}=L-5$.
The total number of all other configurations is finite and does not
depend on $L$, so the second central moment is proportional to $L$
but no to $L^{2}$. Substituting the values from the Table
\ref{tab:2crosd} into \eq{eq:mu2d1}, we arrive at the following
expression for the second central moment $\mu_2^{\rm 1D}$ \be
\mu_2^{\rm 1D}=L \left[\sum \limits_{r=0}^{3} a_{r}^{\rm
1D}J_{r}^{\rm 1D} -\frac{1}{9}L\right]= \frac{2}{45}L. \label{mu2d2}
\ee

To calculate the third central moment $\mu_3^{\rm 1D}$, \be
\mu_{3}^{\rm 1D}=\left< \left(M-\left<M \right> \right)^{3} \right>=
\left< M^{3} \right>-\left< M \right>^{3}-3 \left< M\right> \left(
\left< M^{2} \right>- \left<M \right>^{2} \right), \label{eq:mu3d1}
\ee we proceed exactly along the same lines as above. The averaged
third power of $M$ is \be \left<M^{3} \right> = \int \limits_{0}^{1}
... \int \limits_{0}^{1} \left[\sum \limits_{i=1}^{L} \sum
\limits_{j=1}^{L} \sum \limits_{k=1}^{L}
\Delta_{i}^{-}\Delta_{i}^{+} \Delta_{j}^{-}\Delta_{j}^{+}
\Delta_{k}^{-}\Delta_{k}^{+} \left\{\delta(x_{a}-x_{b}) \right\}
\right] dx_{1}dx_{2}... dx_{L}. \label{eq:M3} \ee This quantity
depends only on integrals over three groups of points
$x_{i-1},x_{i},x_{i+1}$, $x_{j-1},x_{j},x_{j+1}$ and
$x_{k-1},x_{k},x_{k+1}$ and their mutual arrangement. To proceed, we
fix some position of $x_{i}$ (as it has been done for $\mu_{2}^{\rm
1D}$) and consider the positions of $x_{j}$ and $x_{k}$ with respect
to it. Using the diagrammatic approach we compute integrals $G_{r}$
in \eq{eq:M3} for each three--point configuration. The corresponding
diagrams $G_{r}^{\rm 1D}$ for $r=0,\dots,4$ are shown in
Fig.\ref{fig:fig5}b. We encounter the following possibilities:

(i) The integral \eq{eq:M3} contains terms like $G_{0}^{\rm 1D} =\int \int
\theta(x_{i}-x_{j}) \theta(x_{j}-x_{i})dx_{i}dx_{j} =0$;

(ii) All three points $x_{i},x_{j},x_{k}$ are separate. In such a situation the
integrations over three groups of points are independent, giving for each group
$\frac{1}{3}$ according to \eq{eq:mu1d2}. So, we have $G_{1}^{\rm
1D}=[P(\up\down)]^{3}=\frac{1}{27}$. All terms of such a type in \eq{eq:mu3d1}
cancel;

(iii) Two groups have the common points and the third group is separated from them.
In this case the separated integration over the third group of points gives $\la M
\ra$. The integration over the rest pair of groups gives just the same result as the
contribution to the second central moment. The factor 3 in front of $\left( \la
M^{2} \ra- \la M \ra^{2} \right)$ corresponds to three different ways ($ij+k$,
$ik+j$, $jk+i$) to ascribe the indices to these points. Thus, the contributions from
the three--points configurations of such a type and contribution of two--point
configurations for $\la M^{2} \ra$ in \eq{eq:mu3d1} cancel;

(iv) At least two pairs of groups have common points. Such configurations give a
non--zero contribution. Integrals for such groups $G_{r}^{\rm 1d}$ and corresponding
weights $c_{r}^{\rm 1D}$ are summarized in the Table \ref{tab:3crosd} where
$G_{r}=J_{r}$ for $r=0,2,3$.

\begin{table}[ht]
\centerline{\mbox{
\begin{tabular}{|l|c|c|c|c|c|c|}
\hline $r$ & 0 & 1 & 2 & 3 & 4 \\ \hline $G_{r}^{\rm 1D}$ &  0 & $\frac{1}{27}$ &
$\frac{2}{15}$ & $\frac{1}{3}$ & $\frac{17}{315}$ \\ \hline $c_{r}^{\rm 1D}$ & 24 &
$L-37$  & 6 & 1 & 6 \\ \hline
\end{tabular}
\hspace{1cm}
\begin{tabular}{|l|c|c|c|c|} \hline $r$ & 0 & 1 & 2 & 3 \\ \hline $J_r^{\rm 1D}$ & 0
& $\frac{1}{9}$ & $\frac{2}{15}$ & $\frac{1}{3}$ \\ \hline $b_{r}^{\rm 1D}$ & 36 &
18 & 42 & 15 \\ \hline
\end{tabular}
} } \caption{Values of integrals $G_r^{\rm 1D}$ and of the weights $c_{r}^{\rm 1D}$
(left) for the three--point configurations and values of integrals $J_{r}^{\rm 1D}$
and weights $b_{r}^{\rm 1D}$ (right) for computing $\mu_{3}^{\rm 1D}$.}
\label{tab:3crosd}
\end{table}

We can split nontrivial ($r \ne 1$) three--point configuration ($c_{r}^{\rm 1D}$)
into the two--point configuration by deleting of the one--point group in three
different ways. Thus, $3\sum \limits_{r=0 \atop r \ne 1 }^{4}c_{r}^{\rm 1D}=\sum
\limits_{q=0}^{3}b_{q}^{\rm 1D}=111$, where $b_{q}^{\rm 1D}$ is the number of the
two--point configurations of type $q$ obtained as a result of splitting of all
possible three--points configurations. The two--point configurations $b_{r}^{\rm
1D}$ are enumerated in the Table \ref{tab:3crosd}. The contribution of the
two--point configurations are not compensated by an appropriate term from the
three--points configurations, so we have to take it into account manually. So,
substituting expansion of \eq{eq:M3} into \eq{eq:mu3d1}, we obtain
\be
\barr{lll} \mu_{3}^{2D} & = & \disp \la M^{3} \ra - \la M \ra^{3}-3\la M \ra \left(
\la M^{2} \ra- \la M \ra^{2} \right) \medskip \\ & = & \disp L \left[\sum
\limits_{r=0}^{4} \left(c_{r} G_{r}^{\rm 1D}- \frac{1}{27}L \right) - \frac{1}{3}
\sum \limits_{q=0}^{3} b_{q} \left(J_{q}^{}-\frac{1}{9} \right) \right]
\medskip \\ &  = & \disp -
\frac{2}{945}L \simeq  -0.0021164 L.
\end{array}
\label{eq:mu3d2}
\ee

It is worth mentioning that the moments in Eqs.\eq{eq:mu1d2}, \eq{mu2d2} and
\eq{eq:mu3d2}, calculated using the operator technique, coincide with the ones
obtained on the basis of exact combinatorial approach \eq{stem}.

\section{Conditional probability $p(l)$ of two peaks separated by distance $l$}
\label{sect:2:5}

We aim now at evaluating the probability $p(l)$ of having two peaks
separated by a distance $l$, under the condition that there are no
peaks (i.e. sequences $\up\, \down$) on the interval between these
peaks. According to \cite{ov}, this probability is given by \be p(l)
= \int^1_0 dx \sum Q(x) \label{s}, \ee where the sum is taken over
all possible peak--avoiding rise--and--descent patterns of length
$l$ between two peaks, while $Q(x)$ denote the $Q$--polynomials
corresponding to each given configuration (see the explanations
above).

\begin{figure}[ht]
\epsfig{file=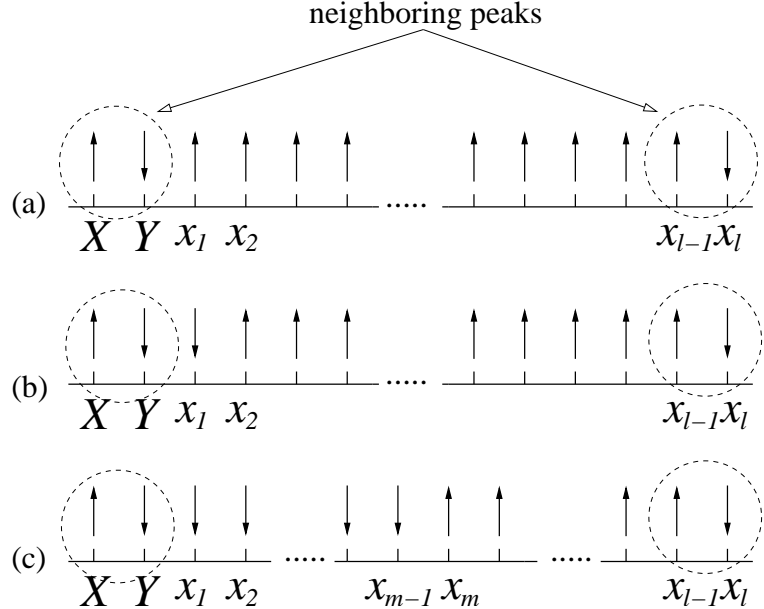,width=10cm} \caption{Rise--and--descent patterns
contributing to the conditional probability of having two closest peaks at distance
$l$ apart from each other. Configuration (a) has only one descent between two peaks.
Configuration (b) has two descents following the first peak, i.e. a "through" at
$x_1$, and (c) presents a generalization of (b) over configurations having $m$
descents, ($m = 1,2, \ldots, l$), after the first peak which are followed by $l-m$
rises, i.e. a "through" at $x_{m-1}$,} \label{f}
\end{figure}

There are several possible peak--avoiding "rise--and--descent"
sequences contributing to such a probability. These sequences are
depicted in \fig{f}. The first peak is located at $0$ position, the
second peak is located at $l$ position. We have fixed a descent at
1st position (variable $Y$) to keep a peak at the 0s position
(variable $X$).

Now, the $Q$-polynomial associated with the sequence (a) in \fig{f}
has the following form: \be Q^a(x) = \int^{1}_x dX \int^{X}_0 dY
\int^1_{Y} dx_1 \int^1_{\disp x_1} dx_2 \ldots \int^1_{\disp
x_{l-2}} dx_{l-1} \int^{\disp x_{l-1}}_0 dx_{l}. \ee Performing the
integration over $x_{l+1}$ in the latter expression, we denote the
iterated integrals over the variables $x_k$, $k =1,2,3 \ldots, l$ as
\be M_l(Y) = \int^1_{Y} dx_1 \int^1_{\disp x_1} dx_2 \ldots
\int^1_{\disp x_{l-1}} x_l \, dx_l. \ee Notice now that $M_l$ obey
the following recursion scheme: \be M_l(Y) = \int^1_{Y}
M_{l-1}(X)dX, \,\,\, M_0(Y) = Y. \ee Introducing the generating
function of the form: \be M(Y) = \sum_{l = 0}^{\infty} M_l(Y) z^l,
\ee we readily find that it obeys \be M(Y) - Y = z \int^1_Y  M(X)dX,
\ee and consequently, $M_l$ are simply the coefficients in the
expansion \be M(Y)=\sum_{l = 0}^{\infty} M_l z^l = \frac{1}{z}
\Big[1 - (1 - z) \exp(z (1 - x))\Big], \ee which are given
explicitly by \be M_l(Y) = \frac{(1-Y)^l}{l!} -
\frac{(1-Y)^{l+1}}{(l+1)!}= \frac{1}{(l+1)!}(l+Y)(1-Y)^{l}. \ee
Hence, the $Q$-polynomial associated with the configuration (a) in
\fig{f} obeys \be Q^a(x) =  \int^1_{x} dX \int^X_0 M_{l-1}(Y)dY=
\frac{1}{l!} \int^1_{x} dX \int^X_0\left[ (1 - Y)^{l-1} (l-1 + Y)
\right] dY, \ee and the contribution of this very configuration to
the probability $p(l)$ reads: \be p^a(l) = \frac{1}{l!} \int^1_0 dx
\int^1_{x} dX \int^X_0 \left[ (1 - Y)^{l-1} (l -1+ Y) \right] dY.
\ee

Next, we turn to the contribution coming out of the general configuration (c) in
\fig{f}. The $Q$-polynomial associated with this configuration of rises and descents
for a fixed $m$ is given by
\beqn
\label{eq:qc} Q^c(x,m) & = &  \int^{1}_x dX \int^{X}_0 dY \int^Y_{0} dx_1
\int^{\disp x_1}_0 dx_2 \int^{\disp x_2}_0 dx_3 \ldots \int^{\disp x_{m-2}}_0
dx_{m-1} \int^1_{\disp x_{m-1}}  dx_{m} \ldots \int^1_{\disp x_{l-2}} dx_{l-1}
\int^{\disp x_{l-1}}_0
dx_{l} \nonumber \\
& = & \int^{1}_x dX \int^{X}_0 dY \int^Y_{0} dx_1 \int^{\disp x_1}_0 dx_2
\int^{\disp x_2}_0 dx_3 \ldots \int^{\disp x_{m-2}}_0 dx_{m-1} \int^1_{\disp
x_{m-1}}  dx_{m} \ldots \int^1_{\disp x_{l-2}} x_{l-1} dx_{l-1} \nonumber \\
&=& \int^{1}_x dX \int^{X}_0 dY \int^Y_{0} dx_1 \int^{\disp x_1}_0 dx_2 \int^{\disp
x_2}_0 dx_3 \ldots \int^{\disp x_{m-2}}_0   M_{l-m}(x_{m-1}) dx_{m-1}.
\eeqn
Let us note, that $Q^{a}(x)=Q^{c}(x,m=1)$. Consider next a recursion scheme of the
form
\be
N_m(Y) = z \int^Y_0 N_{m-1}(X)dX,
\ee
where $N_0(Y)$ is some arbitrary function $\Phi(Y)$. Introducing the generating
function
\be
N(Y) = \sum_{m=0}^{\infty} N_m(Y) z^m,
\ee
we get that it obeys
\be
N(Y) - \Phi(Y) = z \int^Y_0  N(X)dX.
\ee
Solution of the latter equation can be readily obtained by standard means and reads
\be
N(Y) = \Phi(0) \exp(z Y) + \int^Y_0 \frac{d \Phi(X)}{d X} \exp\Big(z (Y - X)\Big)dX.
\ee
Finally, expanding the rhs of the latter equation in powers of $z$, we get that
$N_m(Y)$ are given explicitly by
\be
N_m(Y) = \Phi(0) \frac{Y^m}{m!} + \int^Y_0  \frac{d \Phi(X)}{d X} \frac{(Y -
X)^m}{m!}dX= \int^Y_0  \Phi(X) \frac{(Y - X)^{m-1}}{(m-1)!}dX.
\ee

\begin{figure}[ht]
\epsfig{file=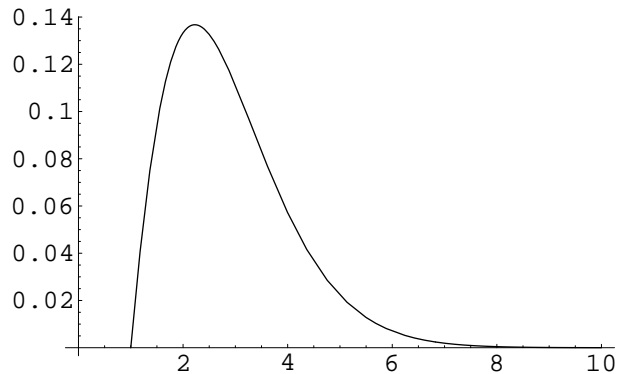, width=9cm} \caption{Probability of having two peaks being
at distance $l$ apart of each other with no peaks inbetween. It has a maximum at
$l=2$.} \label{fig:proba}
\end{figure}

Now, we notice that, as a matter of fact, the multiple integral over the variables
$Y$ and $x_k$, $k = 1,2, \ldots ,m-1$ on the rhs of Eq.(\ref{eq:qc}) becomes just
the function $N_m(Y)$, if one takes $\Phi(Y) = M_{l-m}(Y)$. This implies that \beqn
\int_{0}^{X}d Y \int^Y_{0} dx_1 \int^{\disp x_1}_0 dx_2 ... \int^{\disp x_{m-2}}_0
M_{l-m}(x_{m-1})dx_{m-1} = \int^X_0 \frac{(X-Y)^{m-1} (1-Y)^{l-m} (l - m +Y)}
{(m-1)! (l - m+1 )!}dY. \eeqn Substituting $m=1$ we obtain here $\int
\limits_{0}^{X}M_{l-1} dY$. Consequently, we find that the desired probability obeys
\be
p(l) = \int^1_0 dx \int^1_{x} dX \int^X_0  \left[ \sum_{m = 1}^{l-1}
\frac{(X-Y)^{m-1} (1-Y)^{l-m} (l - m +Y)}{(m - 1)! (l - m+1)!}\right] dY.
\ee
Let us introduce the generating function $F(z)=\sum
\limits_{l=2}^{\infty}z^{l}p(l)$. We can represent the sum
\be
\sum
\limits_{l=2}^{\infty} \sum \limits_{m=1}^{l-1}z^{l}f(l,m) =\sum
\limits_{m=1}^{\infty} \sum \limits_{k=1}^{\infty}z^{k+m}f(k+m,m),
\ee
where $k=l-m$. Using
\be
\sum \limits_{m=1}^{\infty}z^{m}\frac{\left(X-Y\right)^{m-1}}{(m-1)!} =z
e^{z\left(X-Y \right)},
\ee
and
\be
\sum \limits_{k=1}^{\infty} z^{k}\left(k+Y\right) \frac{\left(1-Y
\right)^{k}}{(k+1)!}=e^{z\left(1-Y\right)} \left(1-\frac{1}{z}
\right)+\frac{1}{z}-Y,
\ee
we obtain
\be
\barr{lll} F(z) & = & \disp \sum \limits_{l=2}^{\infty} z^{l} \left( \int^1_0 dx
\int^1_{x} dX \int^X_0 \left[ \sum_{m = 1}^{l-1} \frac{(X-Y)^{m-1} (1-Y)^{l-m} (l -
m +Y)}{(m - 1)! (l - m+1)!}\right] dY \right) \medskip \\ & = & \disp \int^1_0 dx
\int^1_{x} dX \int^X_0 \left[ \sum_{m = 1}^{\infty} \sum_{k = 1}^{\infty} \left(
z^{m}\frac{(X-Y)^{m-1}}{(m - 1)!} \right) \left(z^{k} \frac{ (1-Y)^{k} (k +Y)}{(
(k+1)!} \right) \right] dY
\medskip \\ & = & \disp \int^1_0 dx \int^1_{x} dX \int^X_0 \left[ z
e^{z(X-Y)}\left(e^{z(1-Y)}\left(1-\frac{1}{z} \right) -Y+\frac{1}{z} \right) \right]
dY \medskip \\ & = & \disp \frac{(z-1)^{2}}{2z^{3}} e^{2z}+\left(
\frac{1}{3}+\frac{1}{2z}-\frac{1}{2z^{3}} \right).
\earr
\ee
Hence, the generating function is given by
\be
F(z)= \frac{2}{15}z^{2}+ \frac{1}{9}z^{3}+ \frac{2}{35}z^{4}+ \frac{1}{45}z^{5}+
\frac{4}{567}z^{6}+\cdots.
\ee
Note that the numerical values $\frac{2}{15}$ (for $l=2$) and $\frac{1}{9}$ (for
$l=3$) correspond to the values obtained in \cite{ov}. In general, we get the
following explicit expression
\be
p(l)=\frac{1}{2}\left( \frac{2^{l+1}}{(l+1)!} -2\frac{2^{l+2}}{(l+2)!}
+\frac{2^{l+3}}{(l+3)!} \right)=2^{l}\frac{(l-1)(l+2)}{(l+3)!}, \label{eq:p(l)}
\ee
which defines the probability $p(l)$ of finding two peaks separated by the distance
$l$ with no peaks between these two points for arbitrary $l$. The function $p(l)$ is
depicted in \fig{fig:proba}.

Note also that $p(l)$ in eq.(\ref{eq:p(l)}) coincides with the
distribution function of the distance between two "weak" bonds
obtained by Derrida and Gardner \cite{derrida2} in their analysis of
the number of metastable states in a one--dimensional random Ising
chain at zero's temperature.

\section{Probability distribution function $P(M,L)$ in 2d.}
\label{sect:3}

In this last section we are going to study the statistics of peaks
in a two--dimensional BD landscape $h(i,j,n)$ by converting this
problem to the analysis of the distribution of peaks in the
associated permutation matrix.

It is convenient to represent the two--dimensional base $(i,j)$,
($i,j =1,2,3, \ldots, m$, $m \times m = L$) as a one--dimensional
set with long--ranged correlations. Namely, reading the lattice
$(i,j)$ from the left to the  right in the line and line-by-line
from top--to--bottom, exactly as an electron beam does to highlight
the image on the TV screen, we can rewrite equations \eq{eq:4} and
\eq{eq:5} renaming $h(i,j,n)$ as: $h(i,j,n) \equiv h(k,n)$, where $k
= 1,2,3, \ldots, L$. Hence, $h(i-1,j,n)\equiv h(k-1,n);\,
h(i+1,j,n)\equiv h(k+1,n);\, h(i,j-1,n)\equiv h(k-m,n);\,
h(i,j+1,n)\equiv h(k+m,n)$, where $1\le k\le L$.

Given Eq.\eq{eq:4}, we can describe the growth of a two--dimensional
BD landscape over the base $m\times m$ as a stationary ``updating
dynamics'' in the $L \times L $ permutation matrix with the uniform
distribution on updating events. The uniform ``updating dynamics''
on permutations generates the PDF of peaks identical to the PDF of
peaks computed over the ensemble of all $L!$ equally weighted
permutations (compare to the one--dimensional case). Thus, repeating
the arguments of the previous section, we can construct an analogue
a two-dimensional PGRW for permutation matrix with finite--length
correlations.

The most significant and the only difference between one-- and
two--dimensional models is in the definition of a "peak". In 1D
case, a peak appears at the position $k$ of the permutation matrix
if the corresponding permutation $\pi_k$ is larger then the nearest
neighboring permutations $\pi_{k-1}$ and $\pi_{k+1}$. In 2D case,
$k$ is a peak if and only if $\pi_k$ is larger than $\pi_{k-1},
\pi_{k+1}, \pi_{k-m}$ and $ \pi_{k+m}$ simultaneously.  In the next
subsection we generalize the operator formalism over the 2D case and
calculate the PDF of peaks for the uniform BD process in two
dimensions.

\subsection{Central moments of the probability distribution function in 2D.}

To obtain the limiting ($L\to\infty$) Probability Distribution
Function $P(M,L)$ of having exactly $M$ peaks on a two--dimensional
square base $L=m\times m$, we calculate the first three central
moments of the distribution function and then construct the
corresponding Edgeworth series (cumulant expansion) \cite{edg}. This
enables us: a) to show that in the limit $L\to\infty$ the function
$P(M,L)$ converges to the Gaussian distribution, and b) to present
an  explicit expression for $P(M,L)$ in this limit.

The total number of points for 2D system is $L=m^{2}$. Applying to
the 2D case the same arguments as in 1D case, we define \be
\barr{ll}
\Delta_{i,j}^{\up}=\theta(x_{i,j}-x_{i-1,j}) & \mbox{for the "northern" neighbor} \medskip \\
\Delta_{i,j}^{\rig}=\theta(x_{i,j}-x_{i,j+1}) & \mbox{for the "western" neighbor} \medskip \\
\Delta_{i,j}^{\down}=\theta(x_{i,j}-x_{i+1,j}) & \mbox{for the "southern" neighbor} \medskip \\
\Delta_{i,j}^{\lef}=\theta(x_{i,j}-x_{i,j-1}) & \mbox{for the
"eastern" neighbor} ,\earr \label{eq:Delta2D} \ee (compare
\eq{eq:Delta2D} to \eq{eq:Delta}).

Using the operator formalism we obtain for the first central moment
\be \barr{lll} \mu_{1}^{\rm 2D} & = & \disp \int \limits_{0}^{1} ...
\int \limits_{0}^{1} \left[\sum \limits_{i=1}^{m}\sum
\limits_{j=1}^{m} \Delta_{i,j}^{\up}\Delta_{i,j}^{\rig}
\Delta_{i,j}^{\down}\Delta_{i,j}^{\lef}
\right] dx_{1,1}dx_{1,2}\cdots dx_{m,m-1} dx_{m,m} \medskip \\
& = & \disp L \int \limits_{0}^{1}...\int \limits_{0}^{1}
\Delta_{i,j}^{\up}\Delta_{i,j}^{\rig}
\Delta_{i,j}^{\down}\Delta_{i,j}^{\lef}\; dx_{i-1,j} dx_{i,j+1}
dx_{i+1,j} dx_{i,j-1} dx_{i,j} = \frac{1}{5} L.
\end{array}
\label{eq:m1}
\ee
As in 1D, all terms in \eq{eq:m1} are identical and independent.

\begin{figure}[ht]
\epsfig{file=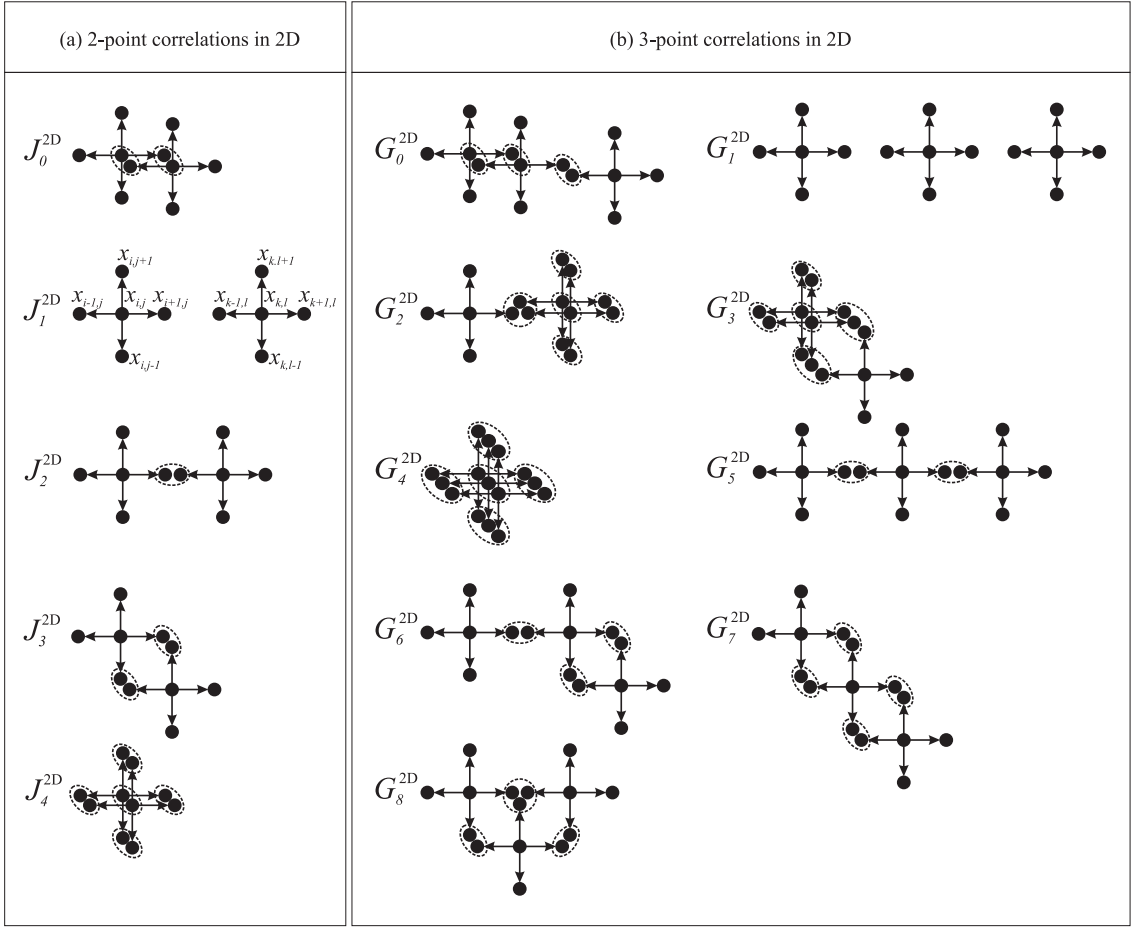,width=15cm} \caption{Basic diagrams and corresponding
integrals: a) $J_{r}^{2D}$ for the computation of $\mu_{2}^{\rm 2D}$; b)
$G_{r}^{2D}$ for the computation of $\mu_{3}^{\rm 2D}$.} \label{fig:fig3}
\end{figure}

2. The same method is used for computing the second central moment. We define $d
{\bf x}_{i,j}=dx_{i-1,j}dx_{i,j+1} dx_{i+1,j}dx_{i,j-1} dx_{i,j}$ for integration
over the point $x_{i,j}$ and its 4 neighbors to make the expressions more compact.
Instead of summing over $i,j$, we fix the position of the first point $x_{i,j}$ and
perform the summation over all possible positions of $x_{k,l}$ with respect to
$x_{i,j}$. Then we enumerate all different integrals $J_{r}^{\rm 2D}$ and compute
the corresponding weights $a_{r}^{\rm 2D}$. The configurations which contribute to
$J_{r}^{\rm 2D}$ are shown in Fig.\ref{fig:fig3}a,b. As in 1D case, the
$\delta$--functions $\delta(x_{a,b}-x_{c,d})$ cut off the diagrams with coinciding
points (see the explanation after \eq{eq:mu2d1}).

The values of integrals $J_{r}^{\rm 2D}$ and their weights $a_{r}^{\rm 2D}$ are
collected in Table \ref{tab:2Dmu2}.

\begin{table}[ht]
\begin{tabular}{|l|c|c|c|c|c|}
\hline $r$ & 0 & 1 & 2 & 3 & 4 \\ \hline $J_{r}^{\rm 2D}$ & 0 &
$\frac{1}{25}$&$\frac{2}{45}$ & $\frac{1}{20}$ & $\frac{1}{5}$ \\ \hline $a_{r}^{\rm
2D}$ & 4 & $N-13$ & 4& 4 &1 \\ \hline
\end{tabular}
\caption{Values of integrals $J_r^{\rm 2D}$ and of the weights $a_{r}^{\rm 2D}$ for
computing $\mu_{2}^{\rm 2D}$.} \label{tab:2Dmu2}
\end{table}

The total number of integrals is $\sum \limits_{r=0}^{4} a_{r}^{\rm
2D}=L=m^2$. The contribution from the integral $J_{1}^{\rm 2D}$ is
exactly canceled by the term $-\frac{1}{25}$ coming from two
independent points: \be \barr{lll} \mu_{2}^{\rm 2D} & = & \disp \int
\limits_{0}^{1} ... \int \limits_{0}^{1} \left[ \sum
\limits_{i=1}^{m}\sum \limits_{j=1}^{m}\sum \limits_{k=1}^{m}\sum
\limits_{l=1}^{m} \Delta_{i,j}^{\up}\Delta_{i,j}^{\rig}
\Delta_{i,j}^{\down} \Delta_{i,j}^{\lef}\;\;
\Delta_{k,l}^{\up}\Delta_{k,l}^{\rig} \Delta_{k,l}^{\down}
\Delta_{k,l}^{\lef}\left\{\delta(x_{a,b}-x_{c,d}) \right\}
\right] dx_{1,1}dx_{2,1}... dx_{m,m}-\frac{1}{25} L^{2} \medskip \\
& = & \disp L \sum \limits_{k=1}^{m}\sum \limits_{l=1}^{m}\left[\int
\limits_{0}^{1} ... \int \limits_{0}^{1}
\Delta_{i,j}^{\up}\Delta_{i,j}^{\rig} \Delta_{i,j}^{\down}
\Delta_{i,j}^{\lef}\;\; \Delta_{k,l}^{\up}\Delta_{k,l}^{\rig}
\Delta_{k,l}^{\down}\Delta_{k,l}^{\lef}\;
\left\{\delta(x_{a,b}-x_{c,d}) \right\}
d {\bf x}_{i,j} d {\bf x}_{k,l} -\frac{1}{25} L\right] \medskip \\
& = & \disp L \sum \limits_{r=0}^{4}\left[a_{r}^{\rm 2D }J_{r}^{\rm
2D} -\frac{1}{25} L \right] = \frac{13}{225} L. \earr
\label{eq:mu21} \ee

3. For computation of the third moment in 2D we use again the same method as in 1D.
Namely, we fix the first point $x_{i,j}$ and enumerate all possible configurations
of three points. Different types of three--points configurations are depicted in
Fig.\ref{fig:fig3}b). Other configurations with the same contribution $G_{r}^{\rm
2D}$ and the same topologies but slightly different conformations are not shown in
the figure. The integrals $G_{r}^{\rm 2D}$ and $J_{r}^{\rm 2D}$ have the same values
for $r=0,2,3,4$. In the configuration $G_{4}^{\rm 2D}$ all three points coincide, so
$G_{4}^{\rm 2D}=\frac{1}{5}$. The most general form of the integral $G_{r}^{\rm 2D}$
is as follows
\be
G_{r}^{\rm 2D}= \int \limits_{0}^{1} ... \int \limits_{0}^{1}
\Delta_{i,j}^{\up}\Delta_{i,j}^{\rig}\Delta_{i,j}^{\down}\Delta_{i,j}^{\lef}\;\;
\Delta_{k,l}^{\up}\Delta_{k,l}^{\rig}\Delta_{k,l}^{\down}\Delta_{k,l}^{\lef}\;\;
\Delta_{m,n}^{\up}\Delta_{m,n}^{\rig}\Delta_{m,n}^{\down}\Delta_{m,n}^{\lef} \left\{
\delta(x_{a,b}-x_{c,d}) \right\} d {\bf x}_{i,j} d {\bf x}_{k,l} d {\bf x}_{m,n},
 \label{eq:int}
\ee
where $d {\bf x}_{i,j}=dx_{i-1,j}dx_{i,j-1}dx_{i+1,j}dx_{i,j+1}dx_{i,j}$ and the
$\delta$--functions in \eq{eq:int} cut off the coinciding points. For example, if in
some configuration the points $x_{i,j+1}$ and $x_{k,l-1}$ coincide, then we include
the function $\delta(x_{k,l-1}-x_{i,j+1})$ etc.

It is possible to simplify integrals of such a type by changing the limits of
integration. For example, $\int \limits_{0}^{1} \int \limits_{0}^{1}
\Delta_{i,j}^{\up} dx_{i,j} dx_{i-1,j} = \int \limits_{0}^{1} x_{i,j} dx_{i,j}$. In
such a way the integral $G_{5}^{\rm 2D}$ can be expressed as follows:
\be
G_{5}^{\rm 2D}=\int \limits_{0}^{1} dx_{i,j}\left[x_{i,j}^{3} \int
\limits_{0}^{x_{i,j}}dx_{i,j+1} \left[ \int \limits_{dx_{i,j+1}}^{1} dx_{i,j+2}
\left[x_{i,j+2}^{2} \int \limits_{0}^{x_{i,j+2}} dx_{i,j+3} \left[ \int
\limits_{dx_{i,j+4}}^{1} x_{i,j+4}^{3}dx_{i,j+4} \right] \right]
\right]\right]=\frac{29}{2925}.
\ee
The values of integrals $G_{0}^{\rm 2D}-G_{8}^{\rm 2D}$ are collected in the Table
\ref{tab:2Dmu3}.

\begin{table}[ht]
\mbox{
\begin{tabular}{|l|c|c|c|c|c|c|c|c|c|}
\hline $r$ & 0 & 1 & 2 & 3 & 4 & 5 & 6 & 7 & 8 \\
\hline $G_r^{\rm 2D}$ & 0 & $\frac{1}{125}$ & $\frac{2}{45}$&$\frac{1}{20}$ &
$\frac{1}{5}$ & $\frac{29}{2925}$ & $\frac{121}{10800}$ & $\frac{7}{550}$ &
$\frac{13}{990}$ \\ \hline $c_{r}^{\rm 2D}$& 168 & $N-313$ & 12 & 12 & 1 & 36 & 48 &
12 & 24 \\ \hline
\end{tabular}
\hspace{1cm}
\begin{tabular}{|l|c|c|c|c|c|}
\hline $r$ & 0 & 1 & 2 & 3 & 4 \\ \hline $J_{r}^{\rm 2D}$ & 0 & $\frac{1}{25}$ &
$\frac{2}{45}$ & $\frac{1}{20}$ & $\frac{1}{5}$ \\ \hline $b_{r}^{\rm 2D}$  & 216 &
216 & 252 & 216 & 39 \\ \hline
\end{tabular}}
\caption{Three--point configurations for computing $\mu_{3}^{\rm
2D}$: integrals $G_r^{\rm 2D}$ and weights $c_{r}^{\rm 2D}$ (left)
and integrals $J_{r}^{\rm 2D}$ and weights $b_{r}^{\rm 2D}$
(right).} \label{tab:2Dmu3}
\end{table}

The symmetry of configurations defines the weights $c_{r}$. For example, the diagram
of the integral $G_{5}^{\rm 2D}$ in Fig.\ref{fig:fig3}b has two orientations along
vertical and horizontal lines. Thus the total weight of the integral $G_{5}^{\rm
2D}$ is $c_{5}^{\rm 2D}=36$.

The values  of $c_{r}$ are given in the Table \ref{tab:2Dmu3}.
Totally there are $\sum \limits_{r=0 \atop r \ne 1}^{8} c_{r}^{\rm
2D}=313$ different nontrivial configurations. We can split each of
the nontrivial three--points configuration as it has been done in 1D
case. There are three ways to do it. Values of $b_{r}$ of the
corresponding two--point configurations are shown in the Table
\ref{tab:2Dmu3} with the total number $\sum \limits_{r=0}^{4}
b_{r}^{\rm 2D}=939$ of such configurations. Now it is possible to
compute the third moment \be \barr{lll} \mu_{3}^{\rm 2D} & = & \disp
L \left[\sum \limits_{r=0}^{8} \left(c_{r}^{\rm 2D} G_{r}^{\rm
2D}-\frac{1}{125} L \right) - \frac{1}{5} \sum \limits_{r=0}^{4}
b_{r}^{\rm 2D}\left( J_{r}^{\rm 2D}-\frac{1}{25} \right) \right]
\medskip \\ & = & \disp \frac{512}{32175} L \simeq 0.015913 L. \earr \label{eq:mu31}
\ee

\subsection{The distribution function $P(M,L)$ in the limit $L \to \infty$.}

\begin{figure}[ht]
\epsfig{file=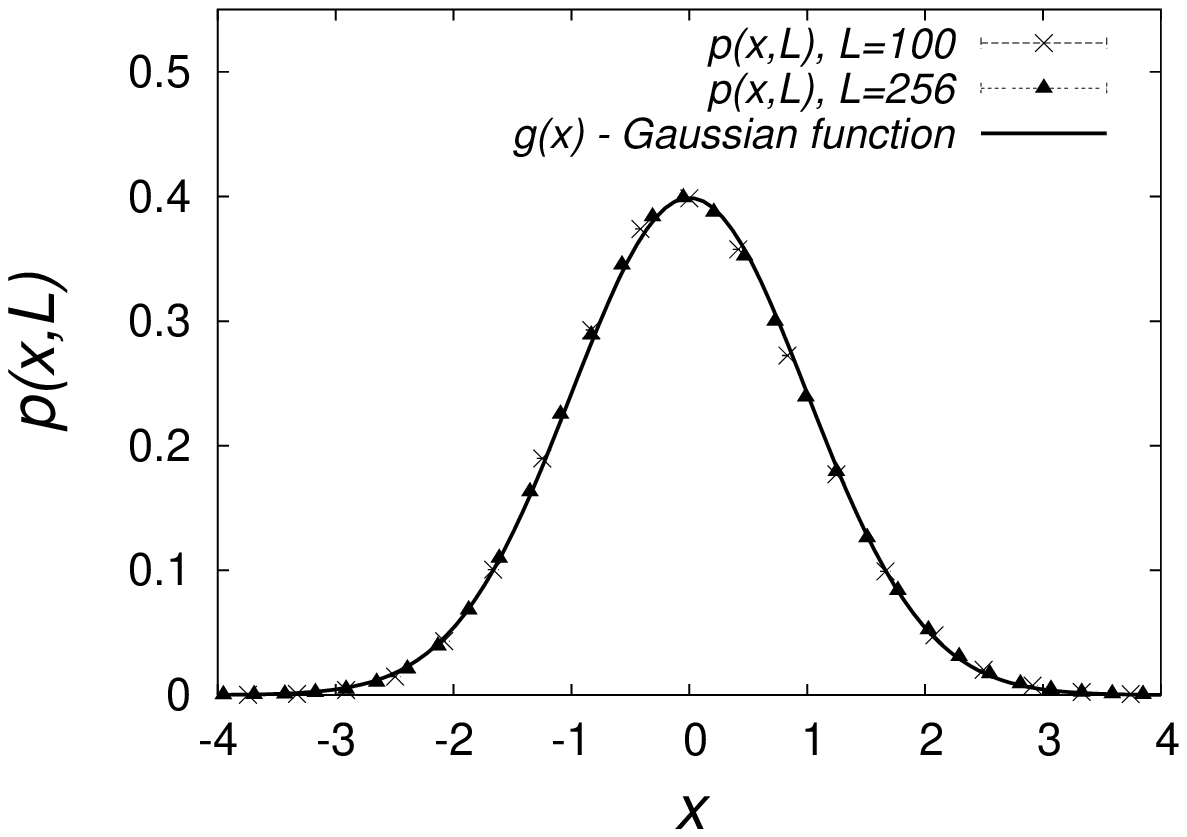,width=8cm} \epsfig{file=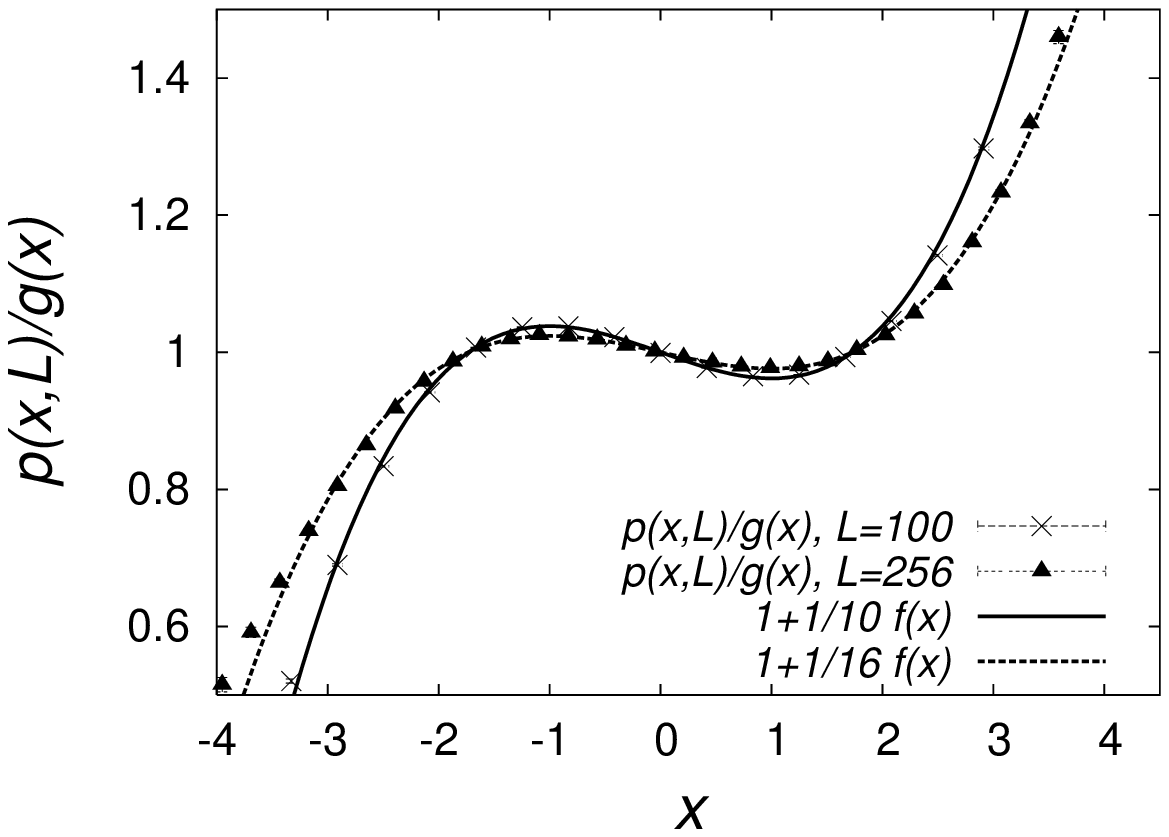,width=8cm}
\caption{(a) Results of numerical simulation of $p(x,L)$ for $L=100, 256$: (a)
comparison of $p(x,L)$ with the Gaussian function $g(x)$; (b) comparison of the
function $p(x,L)/g(x)$ with $1+\frac{1}{\sqrt{L}}f(x)$.} \label{fig:fig7}
\end{figure}

The second and the third cumulants of $P(M,L)$ in 2D are equal to
the second and the third central moments:
\be
\left\{\barr{l} \disp
\kappa_{2}^{\rm 2D}=\mu_{2}^{\rm 2D}=\sigma^{2}= \frac{13}{225}L
\medskip \\ \disp \kappa_{3}^{\rm 2D}=\mu_{3}^{\rm
2D}=\frac{512}{32175} L \earr \right. \label{mom}
\ee
Introducing the normalized deviation, $\disp x=\frac{M-\mu_{1}^{\rm 2D}}{\sigma}$,
we can write the normalized probability distribution $p(x,L)=P\left(\mu_{1}^{\rm
2D}+x \sigma, L \right)$ in a form of the Edgeworth series (cumulant expansion)
\cite{edg}
\be
p(x,L)\simeq g(x) \left(1+\frac{1}{\sqrt{L}}f(x)+
o\left(\frac{1}{\sqrt{L}}\right)\right), \label{cumul0}
\ee
where $g(x)$ is the Gaussian function $\disp g(x)=\frac{1}{\sqrt{2 \pi}}
e^{-x^2/2}$, and $f(x)$ is defined by $\mu_{2}^{\rm 2D}$ and $\mu_{3}^{\rm 2D}$ (see
\cite{edg} for details)
\be
f(x)=\sqrt{L} \frac{\kappa_{3}^{\rm 2D}} {\left(\kappa_{2}^{\rm 2D}
\right)^{3/2}} \frac{1}{6} (x^{3}-3 x) \label{cumul} \ee Substituting \eq{mom} in
\eq{cumul}, we get \be f(x)=\frac{512}{32175} \left(\frac{225}{13} \right)^{3/2}
\frac{1}{6} \left(x^{3}-3 x \right). \label{cumul2}
\ee
Let us emphasize that $f(x)$ does not depend on $L$, as one may
readily check by substituting exact results obtained for the
cumulants to Eq.(\ref{cumul}). To verify the expression in
\eq{cumul2}, we have performed the numerical simulations for the
discrete 2D permutation--generated model with the periodic boundary
conditions and have computed the distribution function $p(x,L)$
numerically. In \fig{fig:fig7}a we present the data of the numerical
simulations for $p(x,L)$ and compare it against the Gaussian
function $g(x)$ for system sizes $L=100$ and $256$. Furthermore, in
\fig{fig:fig7}b we plot the ratio $\frac{p(x,L)}{g(x)}$ as the
function of $x$, which shows that the deviation of the numerically
computed function $p(x,L)$ from the Gaussian function $g(x)$ is
actually very small.

One  clearly sees that for $L\to \infty$ the normalized probability
distribution function $p(x,L)$ converges to the Gaussian function
$g(x)$.

\section{Conclusion}

To conclude, we have studied the probability distribution function $P(M,L)$ of the
number of local peaks in one- and two-dimensional surfaces obtained in a standard
model of next-nearest-neighbor ballistic deposition process.  Our analysis was based
on two central results:

(i) A proof, presented in this paper, of the fact that a ballistic deposition
process in the steady state can be formulated exactly in terms of
``rise--and--descent'' patterns in the ensemble of random permutation matrices,
which made it possible to interpret the BD model in terms of permutations of the set
$1,2,3, \ldots, L$, where $L$ is the number of lattice sites and the numbers drawn
at random from this set determine local heights of the surface.

(ii) An observation made in Ref. \cite{ov} that the ``rise--and--descent'' patterns
can be treated efficiently using a recently proposed algorithm of a Permutation
Generated Random Walk.

In one--dimensional case we have found a closed-form expression \eq{eq:w} for the
generating series $W(s,L)=L! \sum_{M=0}^{\infty} s^{M+1} P(M,L)$ of the distribution
function $P(M,L)$ of peaks in a bounding box of size $L$. Inverting this expression,
we got an exact and asymptotic forms of $P(M,L)$.

Besides, in one-dimension we calculated the probability $p(l)$ of having two peaks
separated by a distance $l$, under the condition that there are no peaks in the
interval between these two peaks. The function $p(l)$ is given by expression
\eq{eq:p(l)}.

For two--dimensional case,  we reformulated the BD process in terms of an "updating
dynamics" on permutations with certain finite-range correlations. Using this
approach, we extended the operator formalism of \cite{ov} to such correlated
permutations and evaluated three first central moments of the PDF. Then, we have
obtained $P(M,L)$ in the asymptotic limit $L\to\infty$ using expansions in the
Edgeworth series \cite{edg}---see \eq{cumul0}--\eq{cumul2}. We have shown that in 2D
the PDF also converges to a Gaussian function as $L \to \infty$.

\begin{acknowledgments}

The authors are grateful to R.Voituriez for helpful discussions. We are also
indebted to B.Derrida for discussions and remarks, as well as for communicating us
his unpublished results. F.H. and S.N. wish to thank the members of the laboratory
LIFR--MIIP (Moscow) for valuable comments and warm hospitality. G.O. acknowledges
financial support from the Alexander von Humboldt Foundation via the Bessel Research
Award. The work is partially supported by the grant ACI-NIM-2004-243 "Nouvelles
Interfaces des Math\'ematiques" (France).

\end{acknowledgments}

\begin{appendix}

\section{}

Let us exploit the relation between the generating function of peaks,
$$
W(s,L)=L! \sum_{M=0}^{\infty} s^{M+1} P(M,L)
$$
and {\em Poly-log} function (see \cite{stem} for details):
\be
\frac{1}{2}\frac{(1+t)^{L+1}}{(1-t)^{L+1}} W\left(\frac{4t}{(1+t)^2},
L\right)=\sum_{m=1}^{\infty} (2m)^{L+1} t^m \label{a1}
\ee
Introducing $s=\frac{4t}{(1+t)^2}$, rewrite \eq{a1} taking into account that
$t=\frac{(1-\sqrt{1-s})^2}{s}$: \be W(s,L)=2 (1-s)^{(L+1)/2}\, \sum_{m=1}^{\infty}
(2m)^{L+1} \left[\frac{(1-\sqrt{1-s})^2}{s}\right]^m \label{a2} \ee

Use now the expansions
\be
\begin{array}{rcl}
\disp (1-s)^{(L+1)/2} & = & \disp 2 \Gamma\left(\frac{L+3}{2}\right)
\sum_{j=0}^{\infty} \frac{(-1)^j}{j!\, \Gamma\left(\frac{L+3}{2}-j\right)}s^j; \medskip \\
\disp (1-\sqrt{1-s})^{q} & = & \disp q\left(\frac{s}{2}\right)^q \sum_{j=0}^{\infty}
\frac{\Gamma(q+2j)}{j!\Gamma(q+j+1)}\left(\frac{s}{4}\right)^j
\end{array}
\label{a3}
\ee

Define now
\be
\begin{array}{l}
\disp a(i,L)= \frac{(-1)^j}{j!\, \Gamma\left(\frac{L+3}{2}-j\right)} \medskip \\
\disp b(m,L) = (2m)^{L+1} 2^{-2m} \medskip \\
\disp c(m,L) = \frac{\Gamma(2m+2j)}{j!\, \Gamma(2m+j+1)}2^{-2j}
\end{array}
\label{a4}
\ee
and rewrite $W(s,L)$ in \eq{a2} as follows
\be
W(s,L) = 2 \Gamma\left(\frac{L+3}{2}\right) \sum_{i=0}^{\infty} a(i,L) s^i
\sum_{m=1}^{\infty} b(m,L) s^m \sum_{j=0}^{\infty} c(m,j) s^j \label{a5}
\ee
After resummation of \eq{a5} we arrive at the following expression:
\be
W(s,L) = 2 \Gamma\left(\frac{L+3}{2}\right) \sum_{M=1}^{\infty} \left[\sum_{l=1}^M
a(M-l,L) \sum_{m=1}^l b(m,L)\, c(m,l-m) \right] s^M \label{a6}
\ee
Hence, by definition of the generating series, we get
\be
\begin{array}{lll}
P(M,L) & = & \disp \frac{2\Gamma\left(\frac{L+3}{2}\right)}{L!} \sum_{l=1}^M
\frac{(-1)^{M-l}}{(M-l)!\,\Gamma\left(\frac{L+3}{2}-M+l\right)} \sum_{m=1}^l
\frac{(2m)^{L+1}}{(l-m)!\,\Gamma(m+l+1)} \medskip \\
& = & \disp \frac{2^{L+2}}{L!} \sum_{l=1}^M (-1)^{M-l} {\frac{L+1}{2} \choose M-l}
\sum_{m=1}^l \frac{m^{L+1}}{(l-m)! (l+m)!}
\end{array}
\label{a7}
\ee
what gives the desired final expression for PDF $P(M,L)$ in one dimension.

\end{appendix}


\begin{thebibliography}{99}
\bibitem{micado} Only the particle of the roof ${\cal T}$ can be removed from the pile
without disturbing the rest of the heap---as in the {\em micado} game.
\end{thebibliography}
\end{document}